\documentclass[runningheads]{llncs}
\usepackage[T1]{fontenc}
\usepackage{graphicx}
\usepackage{cite}
\usepackage[colorlinks=true,
            linkcolor=blue,   % internal links
            citecolor=blue,   % citations
            urlcolor=blue]    % URLs
           {hyperref}
\usepackage{color}

\usepackage{multirow}
\usepackage{graphicx}
\usepackage{subcaption}
\usepackage{float}
\begin{document}
\title{VisG AV-HuBERT: Viseme-Guided AV-HuBERT} %\thanks{}}
% \orcidID{0009-0000-3783-2730} \orcidID{0000-0002-9274-209X}
\author{Aristeidis Papadopoulos, Rishabh Jain \and Naomi Harte}
\authorrunning{A. Papadopoulos et al.}
\institute{Sigmedia, School of Engineering, Trinity College Dublin, Ireland \\
\email{\{papadoar,rijain,nharte\}@tcd.ie}}
\maketitle              % typeset the header of the contribution
\begin{abstract}
% The abstract should briefly summarize the contents of the paper in 150--250 words.
Audio-Visual Speech Recognition (AVSR) systems nowadays integrate Large Language Model (LLM) decoders with transformer-based encoders, achieving state-of-the-art results. However, the relative contributions of improved language modelling versus enhanced audiovisual encoding remain unclear. We propose Viseme-Guided AV-HuBERT (VisG AV-HuBERT), a multi-task fine-tuning framework that incorporates auxiliary viseme classification to strengthen the model's reliance on visual articulatory features. By extending AV-HuBERT with a lightweight viseme prediction sub-network, this method explicitly guides the encoder to preserve visual speech information. Evaluated on LRS3, VisG AV-HuBERT achieves comparable or improved performance over the baseline AV-HuBERT, with notable gains under heavy noise conditions. WER reduces from 13.59\% to 6.60\% (51.4\% relative improvement) at -10 dB Signal-to-Noise Ratio (SNR) for Speech noise. Deeper analysis reveals substantial reductions in substitution errors across noise types, demonstrating improved speech unit discrimination. Evaluation on LRS2 confirms generalization capability. Our results demonstrate that explicit viseme modelling enhances encoder representations, and provides a foundation for enhancing noise-robust AVSR through encoder-level improvements. Code is available at \url{https://github.com/aristosp/visg_avhubert}.

\keywords{Audio-Visual Speech Recognition  \and Multi-task Training \and Visemes}
\end{abstract}
\section{Introduction}
% Bullet - Point List of content.
% \begin{itemize}
%     \item AVSR with LLMs
%     \item Phoneme Viseme information examples
%     \item Multitask learning
%     \item Our method short description
%     \item Structure of paper
% \end{itemize}

Audio-Visual Speech Recognition (AVSR) models have surpassed their audio-only counterparts, especially under noisy conditions where the performance of audio-based models severely degrades. With the advent of Large Language Models (LLMs), transformer-based architectures coupled with LLMs, have reached new state-of-the-art results \cite{llama_avsr,zeroshotavsr,sun2024video-salmonn,mms_llama,cappellazzo25_interspeech,vspllm,vallr}. However, most of these approaches still use transformer-based models such as AV-HuBERT \cite{avhubert} and Auto-AVSR \cite{auto_avsr} as visual feature extractors, and Whisper \cite{whisper} or wav2vec2.0 \cite{wav2vec2} as audio feature extractors. As such, it remains unclear whether these state-of-the-art results reflect improvements in language modelling or in the underlying feature extraction and recognition pipelines.

A less explored direction for improving AVSR currently, is the integration of phoneme or viseme information. Humans rely on visual cues to perceive speech, complementing the auditory signal\cite{visemes}. Similarly, research has shown that models learn phonemes \cite{ma_probing_2021,pasad_layerwise_2021,english23interspeech, seyssel_2022,pasad_comparative_2023} and their visual equivalents, visemes \cite{papadopoulos2025visemes}.
Some efforts have been made to explore the integration of this %type of 
information into the fine-tuning of AVSR models, but the domain remains largely under-explored \cite{univpm, vallr}. The integration of such auxiliary information can be effectively implemented through multi-task learning frameworks. Multi-task learning \cite{caruana1997mtl} enables models to learn more robust shared representations by jointly optimising multiple related tasks, benefiting from inductive transfer between complementary objectives. In AVSR, multi-task learning has already been applied with some promising results\cite{hsu2022uhubert,haliassos2025usr,multiavsr}. Our work takes a fundamentally different approach though, explicitly incorporating visual speech information within the multi-task training paradigm.

In our previous work \cite{papadopoulos2025visemes}, we demonstrated that transformer-based models like AV-HuBERT encode viseme-level information in their learned representations. Building on this finding, we propose Viseme Guided AV-HuBERT (VisG AV-HuBERT), which explicitly leverages viseme representations to enhance encoder capabilities by strengthening the model's reliance on visual articulatory features. Our objective is to improve performance in challenging acoustic environments, where fully exploiting visual information offers a path to better performance than the audio modality alone. As such, our method extends AV-HuBERT with a multi-task supervised fine-tuning framework that incorporates viseme classification as an auxiliary task through a small prediction sub-network. We report improvements in both clean and noisy conditions compared to baseline AV-HuBERT, especially under challenging noise conditions (-10 dB, -5 dB). Analysis of prediction errors reveals that our framework substantially reduces substitution errors, demonstrating improved discrimination at the speech unit level. Additionally, analysis of complex sentences demonstrates that viseme guidance benefits challenging utterances.

The structure of this paper is as follows. In Section \ref{sec:avsr_mtl}, we briefly discuss recent AVSR architectures, multi-task learning approaches, and viseme-to-phoneme mapping. In Section \ref{sec:ourmethod}, we present our method, VisG AV-HuBERT, detailing the architectural design and loss function. Section \ref{sec:implementation} describes our experimental setup, while Section \ref{sec:results} presents our findings. Sections \ref{sec:error_analysis} and \ref{sec:ablation} provide in-depth analysis through error type examination and ablation studies. Finally, Section \ref{sec:conclusion} discusses our findings and proposes directions for future work.

\section{Audio-Visual Speech Recognition Methods} \label{sec:avsr_mtl}
% Bullet - Point List of content.
% \begin{itemize}
%     \item AVSR Models
%     \item Example of AVSR models with phoneme/viseme knowledge
%     \item Multitask with AVSR
%     \item Viseme mapping (table), small discussion about phonemes/visemes?
% \end{itemize}
\subsection{Current AVSR architectures}

Audio-Visual Speech Recognition is the task of generating text transcriptions of speech by jointly leveraging acoustic signals and visual cues from a speaker’s face. Recent AVSR systems have focused on achieving state-of-the-art performance by integrating LLM-based decoders with pre-trained audio and visual encoders. Cappellazzo et al. \cite{llama_avsr} employ Whisper \cite{whisper} and AV-HuBERT \cite{avhubert} for audio and visual feature extraction, with learned modality projectors connecting to Llama family models \cite{grattafiori2024llama3}. They train only the projectors and LoRA modules \cite{lora} while freezing both encoders and the LLM. Critically, their ablation studies reveal that while both encoder and LLM selection impact performance, encoder architecture choices exert particularly strong influence on final WER, with comparable results achievable across different LLM scales when paired with high-quality encoders. 

Thomas et al. \cite{vallr} diverge from the use of AV-HuBERT. Instead, they demonstrate that a Video Transformer, with phoneme prediction as an intermediate representation, can achieve competitive results using a fine-tuned Llama LLM for the final text conversion. While they focus exclusively on Visual Speech Recognition, their phoneme-centric approach addresses visual ambiguities by reducing the prediction space to linguistically meaningful units. However, their reliance on an LLM decoder makes it difficult to isolate the contribution of phonemic representations from language modelling improvements.

Complementary to these LLM-based approaches, Hu et al. \cite{univpm}, propose a universal viseme-phoneme mapping, based on clustering and AV-HuBERT. They set the number of classes for both phonemes and visemes to be 40, following the number of English phonemes. However, as they also acknowledge, this is not accurate, as the viseme-to-phoneme mapping is one-to-many \cite{bear2017decodingvisemesimprovingmachine,cappelletta2012phonemetoviseme}.

These findings collectively suggest that the path forward for AVSR may not lie in increasingly sophisticated language model decoders, but rather in improving the quality of audiovisual encoder representations. Recent work by Guan et al. \cite{guan2025mllm} on multimodal LLMs reinforces this perspective, demonstrating that developing more powerful visual encoders yields greater performance gains than increasing decoder capacity. Specifically, incorporating explicit phoneme and viseme information into pre-trained audiovisual encoders could enable models to learn features that better capture the relationship between articulatory movements and acoustic signals, while naturally handling the inherent visual ambiguities in speech. Building on this idea, we propose integrating viseme information directly into the encoder. This motivates our focus on enhancing AV-HuBERT through an auxiliary viseme prediction task, rather than adding LLM-based decoders.

\subsection{Multi-task training in AVSR}

Multi-task learning \cite{caruana1997mtl} provides an established framework for incorporating auxiliary objectives into encoder training. Multi-task learning itself is not a new concept, with the core principle being that models can learn more robust and generalizable representations by jointly optimising multiple related tasks, leveraging shared knowledge to improve performance on each individual task. Hsu et al. \cite{hsu2022uhubert} introduced u-HuBERT, extending AV-HuBERT through pre-training on AVSR, ASR, and VSR tasks before task-specific fine-tuning on ASR. However, all tasks are still optimised only at the token level, without explicit constraints on fine-grained visual articulatory structure, which can limit robustness under severe audio corruption or visual-only conditions. Haliassos et al. \cite{haliassos2025usr} employed semi-supervised multi-task learning with student-teacher frameworks, but this approach is sensitive to the quality of pseudo-labels and can be prone to overfitting in low-resource regimes. Most recently, Torrie et al. \cite{multiavsr} proposed a supervised multi-task framework based on Auto-AVSR \cite{auto_avsr} for jointly training on VSR, AVSR, and ASR tasks. This design, however, relies heavily on large amounts of labelled data and again supervises only word/character outputs, providing limited guidance for learning low-level articulatory patterns. By incorporating viseme classification as an auxiliary task, we leverage the multi-task learning paradigm specifically to strengthen the model's reliance on visual articulatory information, directly addressing these limitations.

\subsection{Viseme to phoneme mapping}
Visemes are the visual equivalents of phonemes, the basic speech unit. For English, multiple viseme-to-phoneme mappings have been proposed due to the lack of standardised viseme definitions, as visemes exhibit greater ambiguity than their phoneme counterparts 
\cite{jeffers,bear2017decodingvisemesimprovingmachine,hazen,leemap}. Most of the mappings have the same consonant groupings, but differ in the vowel groupings. As noted in our prior work \cite{papadopoulos2025visemes}, Lee’s mapping \cite{leemap} provides balanced consonant and vowel groupings. We therefore adopt this mapping, shown in Table \ref{tab:lee_table}, using ARPABET notation.

\begin{table}[htbp]
\caption{Lee Phoneme to Viseme Mapping}
\begin{center}
\scalebox{0.9}{
\begin{tabular}{|c|c|c|c|}
\hline
\textbf{Viseme} & \textbf{Phonemes} & \textbf{Viseme} & \textbf{Phonemes} \\
\hline
$F$   & \textit{F V} & $IY$  & \textit{IY IH} \\
$W$   & \textit{R W} & $EH$  & \textit{EH EY AE} \\
$P$   & \textit{B P M} & $AA$  & \textit{AA AW AY} \\
$K$   & \textit{G K NG N L Y HH} & $AH$  & \textit{AH} \\
$T$   & \textit{T D S Z DH TH} & $AO$  & \textit{AO OY OW} \\
$CH$  & \textit{CH JH SH ZH} & $UH$  & \textit{UH UW} \\
$S$ & \textit{/$sil$/} & $ER$  & \textit{ER} \\
\hline
\end{tabular}
}
\label{tab:lee_table}
\end{center}
\end{table}

\section{Viseme-Guided Audio-Visual Speech Recognition}\label{sec:ourmethod}
% In this section, we go into details regarding our proposed model, VisG AV-HuBERT.
\subsection{AV-HuBERT}
AV-HuBERT\cite{avhubert} learns from unlabelled audio-visual data in a self-supervised training. The model comprises of four components: a visual feature extractor, based on a modified ResNet, an audio feature extractor which is a simple feed-forward network (FFN), a fusion module, and a Transformer-based back-end. The two modality front-ends extract frame-level representations, which are concatenated by the fusion module to create the audio-visual features. These features are then provided to the transformer layers, which produce contextualized audio-visual embeddings. Pre-training consists of the independent masking of the two modalities, with the task being to identify the fake frames and then find the original labels. The model can then be fine-tuned for audio-visual or visual speech recognition tasks.

The authors \cite{avhubert} provide two configurations: Base and Large. In the Base model, the encoder comprises 12 Transformer blocks with an embedding dimension of 768, a feed‑forward dimension of 3072, and 12 attention heads. In the Large configuration, the encoder comprises 24 Transformer blocks, with an embedding dimension of 1024, a feed‑forward dimension of 4096, and 16 attention heads.

\subsection{Viseme-Guided AV-HuBERT}
\begin{figure}[htbp]
\centering
\includegraphics[width=0.9\textwidth]{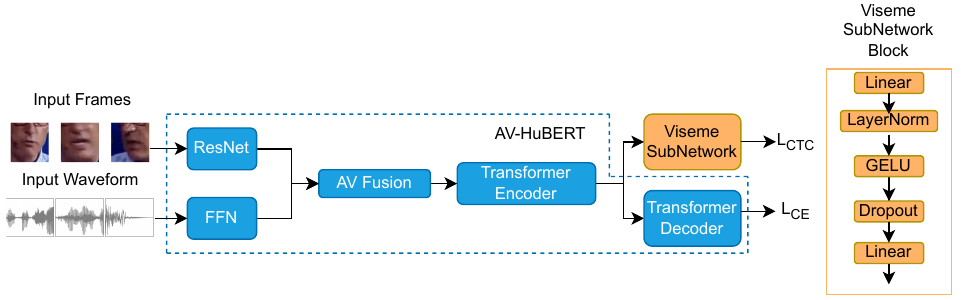}
\caption{VisG AV-HuBERT architecture showing the original AV-HuBERT (dotted outline) and the auxiliary viseme prediction sub-network (solid outline) added during fine-tuning.} \label{model_architecture}
\end{figure}

Our framework extends AV-HuBERT with a lightweight auxiliary sub-network for viseme classification, illustrated in Fig. \ref{model_architecture}. This sub-network, attached to the encoder output, consists of a linear projection layer, layer normalization, GELU activation function, and dropout (p=0.3), followed by a final output layer that predicts 14 + 1 (blank token for CTC loss) viseme classes. By incorporating viseme classification as a secondary task during fine-tuning, the model is encouraged to discover and preserve visual articulatory features that distinguish viseme classes. For each frame with an assigned viseme label, the framework jointly optimizes both the primary AVSR objective and the auxiliary viseme classification task, using the hybrid CTC/CE loss described in Section \ref{sec:loss}.

\subsection{Loss Function} \label{sec:loss}
Suppose \(\mathbf{x} = [x_1, \dots, x_T]\) and \(\mathbf{y} = [y_1, \dots, y_T]\) are the fused feature sequence and the target labels, respectively, with \(T\) being the sequence length. The original framework uses attention-based sequence-to-sequence Cross-Entropy (CE) loss \cite{bahdanau2016attention} during fine-tuning, which is defined as
\begin{equation} \label{eq:1}
    L_{CE} =  \sum_{i} \log p_{CE}(\mathbf{y} \mid \mathbf{x}^i),
\end{equation}
where \(p_{CE}\) is a direct estimation of the posterior using the chain rule, i.e., \(p_{CE}(\mathbf{y} \mid \mathbf{x}) \approx \prod_{t=1}^{T} p(y_t \mid \mathbf{y}_{<t}, \mathbf{x})\), and \(i\) indexes training samples in the batch.

For the auxiliary viseme classification task, we adopt the Connectionist Temporal Classification (CTC) loss \cite{ctc_loss}, which assumes conditional independence between each output prediction: \(p_{CTC}(\mathbf{y} \mid \mathbf{x}) \approx \prod_{t=1}^{T} p(y_t \mid \mathbf{x})\), and is defined as
\begin{equation}\label{eq:2}
    L_{CTC} = \sum_{i} \log p_{CTC}(\mathbf{y} \mid \mathbf{x}^i).
\end{equation}
% Add why
CTC is particularly well-suited for frame-level auxiliary tasks like viseme classification because its blank token mechanism naturally handles cases where multiple consecutive frames map to a single viseme label. We provide an example in Section \ref{sec:dataset}.
We combine these losses using a hybrid CTC/CE framework \cite{hybrid_ctc_ce}:
\begin{equation}\label{eq:3}
    L = L_{CE} + \alpha \, L_{CTC},
\end{equation}
where \(\alpha\) controls the relative weight of the auxiliary viseme task loss. Through grid search, we determined optimal values of \(\alpha = 0.2\) for the Base model and \(\alpha = 0.15\) for the Large model. Details of the grid search procedure are provided in Section \ref{sec:ablation}. This hybrid approach balances complementary strengths: the CE loss provides autoregressive modelling with attention mechanisms and removes conditional independence assumptions, while the CTC loss enforces monotonic alignment and naturally suits frame-level auxiliary tasks. The smaller \(\alpha\) value for the Large model reflects its greater capacity to learn the primary AVSR objective without requiring strong regularization from the auxiliary task.

\section{Experimental Setup} \label{sec:implementation}
% List of content.
% \textbf{Discuss Pre-processing, LRS3, cite AV-HuBERT, MUSAN, Montreal Forced Aligner, Model variants (Large/Base), difference in layers, epochs, noise probability and dB.}
\subsection{Datasets} \label{sec:dataset}
\begin{figure}[htbp]
\centering
\includegraphics[width=\textwidth]{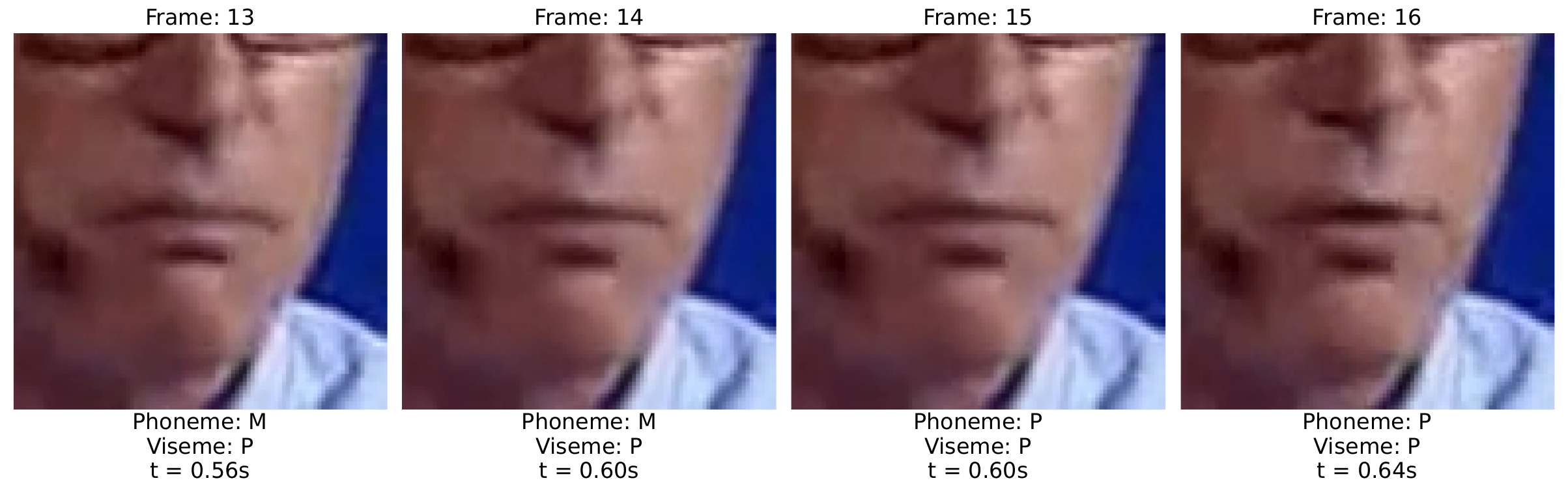}
\caption{Example of frame-level viseme label assignment with multiple consecutive frames mapped to a single viseme class.}
\label{example}
\end{figure}

In our fine-tuning experiments, we use LRS3 \cite{lrs3}, the largest publicly available audio-visual speech recognition dataset with transcriptions. Videos have a resolution of 224 × 224 pixels at 25 FPS. We preprocess the dataset using the AV-HuBERT pipeline \cite{avhubert}, resulting in a low-resource split of 30 hours and the full 433-hour dataset. We train the Base model on the low-resource setting and the Large variant on the full dataset. Furthermore, we follow \cite{noisy_avhubert} to augment audio files with babble, music and random noise samples from the MUSAN dataset \cite{MUSAN} and speech noise from LRS3, ensuring no speaker overlap. Speech noise differs from babble in that it mixes the target speech with a single overlapping speaker, whereas babble combines multiple concurrent speakers. For generalization assessment, we additionally test on the LRS2 test set \cite{LRS2}, an out-of-domain dataset of BBC television clips. We pre-process LRS2 with the same pipeline.
% \cite{gorman2011prosodylab}
To generate the viseme labels, we use Montreal Forced Aligner (MFA) \cite{mcauliffe17_interspeech} with a general American English dictionary. In particular, we first obtain the alignments using the audio files and their corresponding transcriptions. Then, we map the phonemes to Lee's mapping, as shown in Table \ref{tab:lee_table}. Finally, using the alignment timestamps, we map the frames to their respective viseme label, as shown in Figure \ref{example}. We include a blank label for the CTC loss.

\subsection{Implementation Details}
We conduct experiments using both the Base and Large AV-HuBERT variants. For the Base model, the viseme prediction subnetwork consists of a linear layer, layer normalization, GELU activation, and dropout (p=0.3), followed by an output layer projecting to 15 viseme classes. For the Large variant an identical intermediate block is added before the final output layer. 

We fine-tune both models with different noise hyper-parameters, compared to AV-HuBERT, with noise probability p=0.3 (increased from p=0.25) and SNR of -2.5dB (decreased from 0dB). The Base model is trained for 30K updates, with the encoder unfrozen at 20K updates when the viseme classification task is introduced. To smoothly integrate this auxiliary task, we apply a 4K-update warm-up period with linearly increasing loss weight. The Large model is trained for 60K updates with the encoder unfrozen at 40K updates, and in this case the viseme classification loss is applied from the beginning without warm-up.%For the large model, we introduce the viseme task from the beginning without warm-up, as preliminary experiments showed the larger capacity handled immediate multi-task learning effectively.

\section{Performance Evaluation} \label{sec:results}
In this section, we present the results from our experiments. We evaluate VisG AV-HuBERT against baseline AV-HuBERT \cite{avhubert} across clean and noisy conditions. We omit comparisons with LLM-based AVSR systems as our method targets encoder improvements rather than decoder architectures. To assess the significance of performance differences, we perform paired-sample t-tests comparing per-utterance WER between VisG AV-HuBERT and baseline AV-HuBERT. We report statistical significance at the $p$ < 0.05 level, based on paired t-test, with significance indicators (* for $p$ < 0.05, ** for $p$ < 0.01) provided in result tables. % where applicable.
% List of content for this section
% \textbf{Discuss improvement in clean and mainly under 0dB. Highlight speech noise improvement but make note of the babble noise underperformance.}
% \subsection{Results under Clean Conditions}
In Table \ref{tab:clean_wer}, we present the WER results for clean speech, for the LRS3 and LRS2 datasets. VisG AV-HuBERT Base achieves WER reductions of 0.32\% in LRS3 and slightly degrades in performance in LRS2, while VisG AV-HuBERT Large marginally improves in LRS3 and reduces the WER by 0.375\% in LRS2, compared to baseline AV-HuBERT. %, as seen in Table \ref{tab:clean_wer}.

\begin{table}[htbp]
\centering
\caption{WER (\%) Comparison under Clean Conditions}\label{tab:clean_wer}
\scalebox{0.95}{
\begin{tabular}{|l|l|c|c|}
\hline
Model &  Params & WER LRS3 $\downarrow$ & WER LRS2 $\downarrow$\\
\hline
AV-HuBERT Base & 160M & 4.1 & \textbf{10.09}\\
VisG AV-HuBERT Base & 162M & {\bfseries 3.78$^{*}$} & 10.58\\
\hline
AV-HuBERT Large & 477M & 1.4 & 10.3\\
VisG AV-HuBERT Large & 484M & {\bfseries 1.38} & \textbf{9.925} \\
% Category for base/large 
\hline
\end{tabular}
}
\end{table}

Improvements in clean conditions are modest, as in clean acoustic conditions the predictions are based predominantly on the acoustic signal \cite{stewart2014robust,lim2025improving,lin2025uncovering}. %However, the viseme-guided framework demonstrates substantially larger gains under noisy scenarios, particularly at severe SNR levels (-10 dB, -5 dB) where explicit visual articulatory guidance becomes increasingly valuable for disambiguation, as discussed in the next section.

% \subsection{Results under Noisy Conditions}
In Table \ref{tab:wer_noise}, we present the results in noisy conditions for the LRS3 Dataset, ranging from {-10dB to 10dB}, for Babble, Speech, Music and Random Noise types. % 
\begin{table}[htbp]
\centering
\caption{WER (\%) Comparison under Different Noise Types for LRS3}
\label{tab:wer_noise}
\scalebox{0.9}{
\begin{tabular}{|l|c|ccccc|}
\hline
\textbf{Model} & \textbf{Noise Type} & \textbf{-10 dB} & \textbf{-5 dB} & \textbf{0 dB} & \textbf{5 dB} & \textbf{10 dB} \\
\hline
AV-HuBERT Base & \multirow{4}{*}{\centering Babble} & 41.45 & 23.86 & 10.89 & 5.97 & 5.00 \\
VisG AV-HuBERT Base &  & 42.55 & \textbf{22.60$^{*}$} & \textbf{10.18$^{*}$} & 6.03 & \textbf{4.65$^{*}$} \\
\cline{1-1}\cline{3-7}
AV-HuBERT Large &  & 30.00 & 13.45 & 4.78 & 2.52 & 1.91 \\
VisG AV-HuBERT Large &  & 31.03 & 14.05 & 4.92 & \textbf{2.51} & \textbf{1.87} \\
\hline
AV-HuBERT Base & \multirow{4}{*}{\centering Speech} & 15.06 & 9.80 & 6.74 & 5.39 & 4.74 \\
VisG AV-HuBERT Base &  & \textbf{13.55$^{**}$} & \textbf{8.78$^{*}$} & \textbf{6.60} & \textbf{5.33} & \textbf{4.66} \\
\cline{1-1}\cline{3-7}
AV-HuBERT Large &  & 13.59 & 4.70 & 3.17 & 1.98 & 1.67 \\
VisG AV-HuBERT Large &  & \textbf{6.60$^{**}$} & \textbf{3.83$^{**}$} & \textbf{2.50$^{**}$} & \textbf{1.94} & \textbf{1.61} \\
\hline
AV-HuBERT Base & \multirow{4}{*}{\centering Music} & 17.36 & 10.01 & 6.74 & 5.08 & 4.50 \\
VisG AV-HuBERT Base &  & \textbf{16.60} & \textbf{9.36} & \textbf{6.12$^{*}$} & \textbf{5.00} & \textbf{4.30} \\
\cline{1-1}\cline{3-7}
AV-HuBERT Large &  & 10.23 & 4.83 & 2.71 & 1.93 & 1.72 \\
VisG AV-HuBERT Large &  & \textbf{9.62} & \textbf{4.63} & \textbf{2.52} & \textbf{1.92} & \textbf{1.61} \\
\hline
AV-HuBERT Base & \multirow{4}{*}{\centering Noise} & 15.44 & 9.03 & 6.65 & 5.36 & 4.51 \\
VisG AV-HuBERT Base &  & \textbf{14.72} & \textbf{8.87} & \textbf{5.59$^{*}$} & \textbf{5.27$^{*}$} & 4.66 \\
\cline{1-1}\cline{3-7}
AV-HuBERT Large &  & 9.82 & 4.88 & 2.55 & 1.87 & 1.72 \\
VisG AV-HuBERT Large &  & \textbf{8.59} & \textbf{4.18} & \textbf{2.29} & 1.94 & \textbf{1.68} \\
\hline
\end{tabular}
}
\end{table}

From Table \ref{tab:wer_noise}, VisG AV-HuBERT outperforms, or is at least on par with baseline AV-HuBERT, across nearly all noise types and SNR levels. The most substantial improvements occur in the Speech noise condition, where visual articulatory cues are most discriminative for separating target speech from background speech.

For the Base model, statistically significant gains are observed in Speech noise at -10 dB (15.06\% to 13.55\%) and -5 dB (9.80\% to 8.78\%). Moderate but consistent improvements are also observed for Music and Random Noise conditions, with VisG AV-HuBERT Base outperforming the baseline at most SNR levels. In Music noise, reductions are seen at 0 dB (6.74\% to 6.12\%), while in Random Noise, improvements occur at 0 dB (6.65\% to 5.59\%) and 5 dB (5.36\% to 5.27\%).

\begin{figure}[htbp]
\centering
\includegraphics[width=0.85\textwidth]{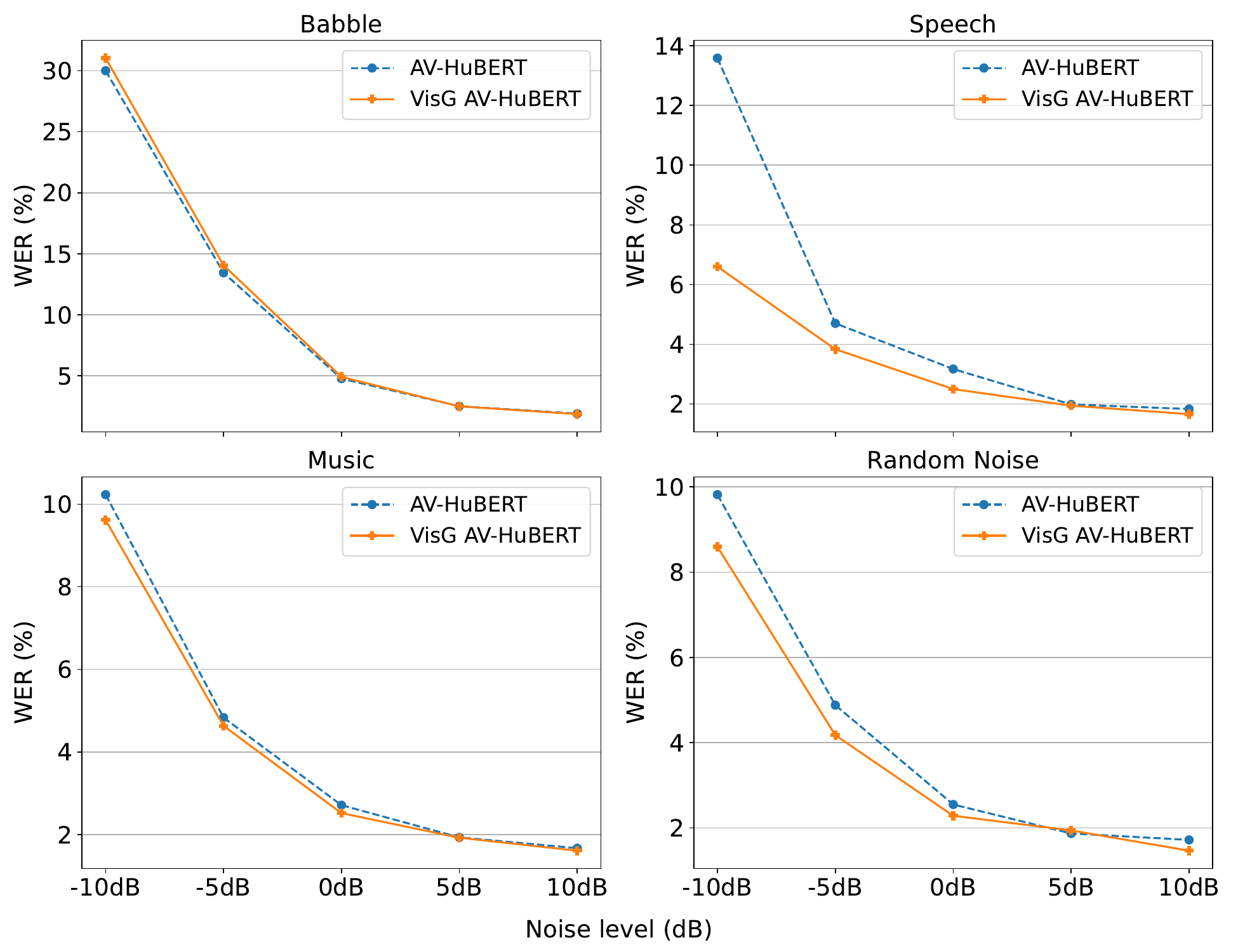}
\caption{LRS3 WER comparison under different noise conditions for Large model.}
\label{fig:wer_noise_comp}
\end{figure}

For the Large model, the improvements are even more pronounced. At -10 dB SNR in Speech noise, VisG AV-HuBERT reduces WER from 13.59\% to 6.60\%, representing a relative improvement of 51.4\%. This improvement remains consistent at -5 dB (4.70\% to 3.83\%, 18.5\% relative reduction) and 0 dB (3.17\% to 2.50\%, 21.1\% relative reduction). Similar gains are observed in Music and Random Noise conditions, with VisG AV-HuBERT Large consistently outperforming the baseline across nearly all SNR levels. A visual inspection of the model's behaviour under noise can be seen in Figure \ref{fig:wer_noise_comp}, which illustrates the performance trends across different noise types and SNR levels for the Large model.

The only exception occurs under Babble noise at extreme SNR levels, where VisG AV-HuBERT Base shows slightly higher WER at -10 dB (42.55\% vs. 41.45\%), though improvements are observed at higher SNRs. The Large model similarly exhibits elevated WERs from -10 dB to 0 dB. This suggests that additional modelling strategies may be needed to handle Babble noise conditions. %  multi-talker interference in extreme Babble noise conditions.

The overall results demonstrate that explicit viseme modelling provides the strongest benefits when discriminating speech from background speech, where visual articulatory information is most informative. The auxiliary viseme task appears to reduce substitution errors by constraining the model to learn discriminative articulatory features that resolve phonetic confusion, as analysed in detail in Section \ref{sec:error_types}. Similar trends are observed on LRS2 under noisy conditions, with further details provided in the Supplementary Material. % and additional results

\section{Error Analysis} \label{sec:error_analysis}
% Bullet - Point List of content.
% \begin{itemize}
%     \item Discuss, show S/I/D/ errors
%     \item Complexity figure  and discussion
% \end{itemize}

% Having established overall performance improvements in the previous section, we now conduct detailed analyses from two perspectives: error type distributions (substitutions, insertions, deletions) and performance across varying sentence complexity levels.

\subsection{Error Types} \label{sec:error_types}
\begin{figure}[htbp]
\centering
\includegraphics[width=0.88\textwidth]{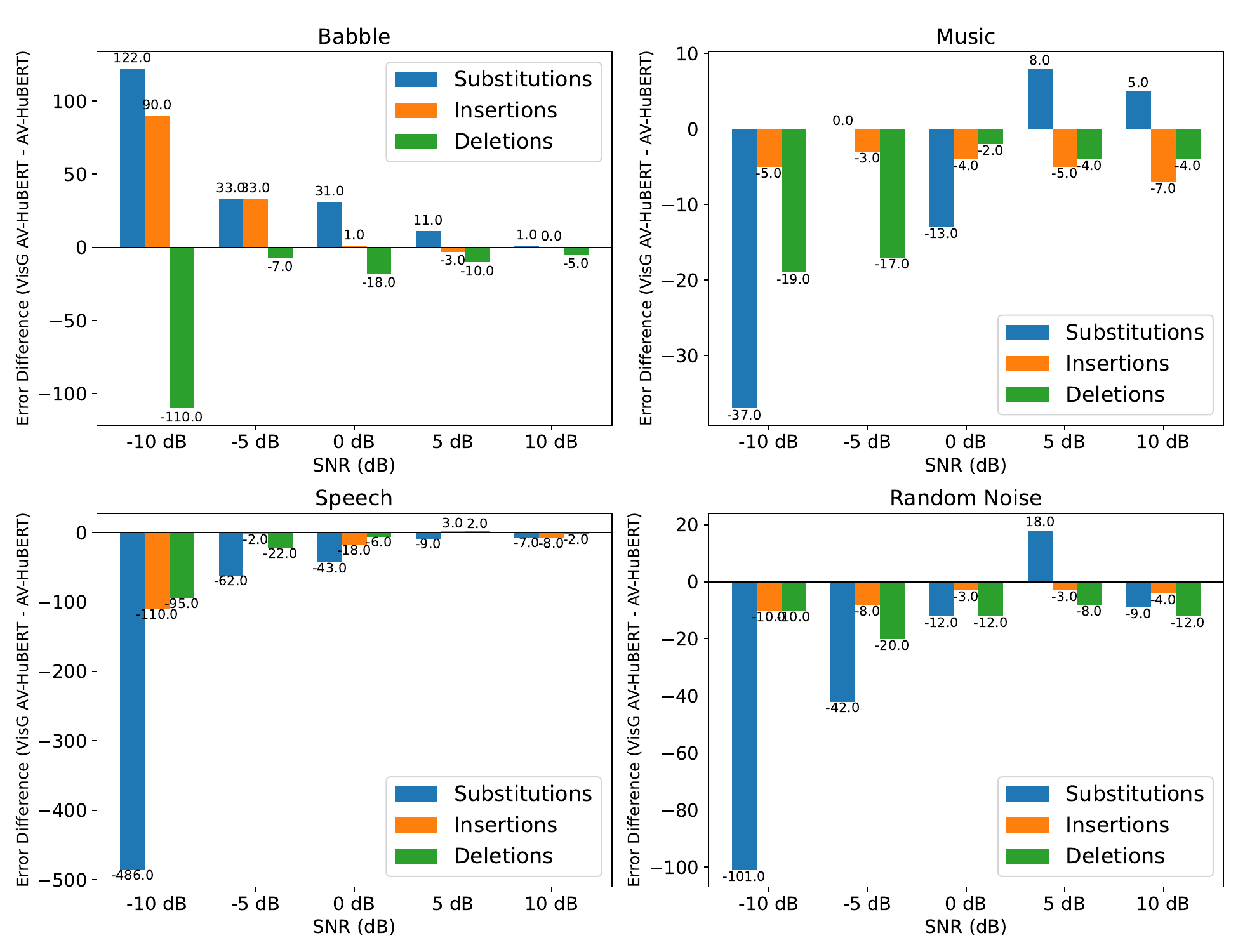}
\caption{Error difference (VisG AV-HuBERT WER minus baseline WER) by error type across noise conditions and SNR levels (Large model). }
 \label{sids_large}
\end{figure}

For each utterance in the LRS3 test set, we compute substitution, insertion, and deletion errors from both VisG AV-HuBERT and baseline AV-HuBERT. We then calculate the difference (VisG minus baseline) in absolute counts for each error category across all SNR levels and noise types, as seen in Figure \ref{sids_large}.

Figure \ref{sids_large} reveals substantial reductions in substitution errors across Speech, Music, and Noise categories, with the largest improvements at -10 dB. Furthermore, we notice reductions in deletion errors which indicates that visual articulatory cues help the model better detect speech unit boundaries, preventing missed detections when acoustic signals are degraded. The auxiliary viseme task successfully guides the model to leverage visual articulatory information alongside audio, thereby reducing phonetic confusion and improving discrimination between similar-sounding phonemes. The consistent pattern across noise types (excluding Babble) demonstrates robust generalization of the viseme-guided approach, validating that visual articulatory information provides guidance regardless of acoustic noise characteristics. Additional error analysis for the Base model can be found in the Supplementary Material.

\subsection{Sentence Complexity}\label{sentence_complexity}
To further demonstrate the successful integration of viseme information in the predictions of the proposed model, we proceed to analyse the difference between the two models in terms of sentence complexity. Following Sterpu et al. \cite{sterpu2020avalign}, we train a simple 5-gram Laplace Language Model on the transcriptions of the LRS3 train and validation set and compute the cross-entropy for each sentence in the test set. In Table \ref{tab:complex_sentences} we provide examples of sentences, sorted by their CE score. Higher CE score means the sentence is more complex.

\begin{table}[htpb]
\centering
\caption{Test set Utterances sorted by their CE Score, with predictions from the Large configurations for each model, under -10dB Speech Noise}\label{tab:complex_sentences}
% \caption{Test Set Utterances sorted by their CE Score, with predictions from the large configurations for each model, under -10dB Speech Noise}\label{tab:complex_sentences}
\scalebox{0.95}{
\begin{tabular}{|l|l|l|l|}
\hline
\textbf{Ground Truth} &  \textbf{CE} & \textbf{AV-HuBERT} & \textbf{VisG AV-HuBERT}\\
\hline
gov just six people & 3.96 & of just six people & of just six people \\
% \hline
% so was mother teresa & 3.62 & so what there's a line of research & so what munder derwica \\
% \hline
% and this leads to oddities & 3.33 & and this leads to honories & and this leads to oddities \\
\hline
the board of ed & 3.15 & the board of it & the boy there \\
\hline
where do refugee hearts go & 2.46 & where did refugee haunts go & where do refugee hearts go \\
\hline
not the wife not the kids & 2.2 & not the wives not the kids & not the wife not the kids \\
\hline
talk to farmers & 1.99 & talk to flambers & talk to farmers \\
\hline
you want to work for him & 1.55 & why don't work for him & you want to work for him \\
\hline
and you know what & 1.3 & i don't know what & and you know what \\
\hline
\end{tabular}
}
\end{table}

\begin{figure}[htpb]
\centering
\includegraphics[width=0.66\textwidth]{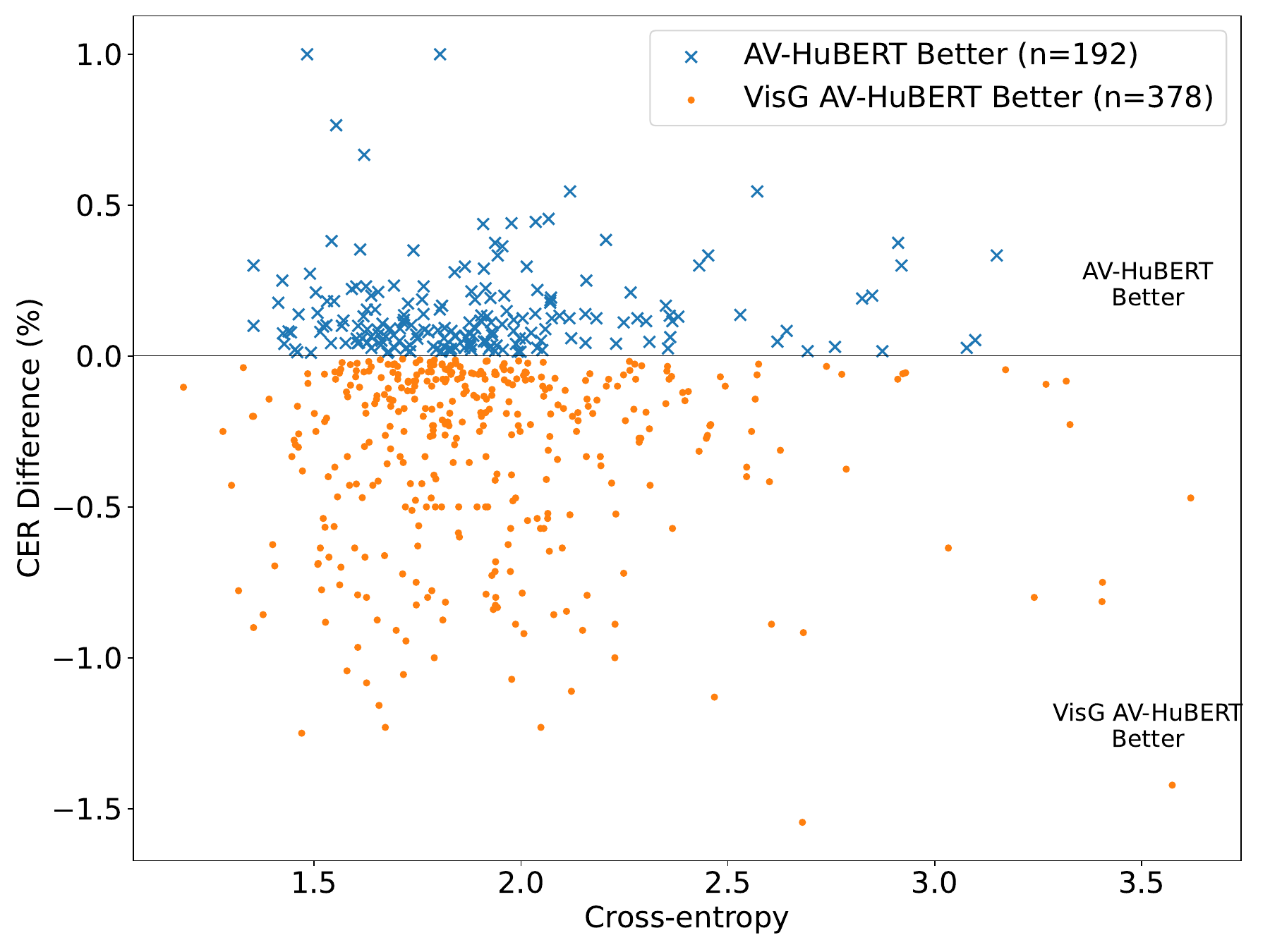}
\caption{CER Difference between VisG AV-HuBERT and AV-HuBERT (Large).}%, under Speech Noise mixed at -10dB.} %  sorted by their complexity
\label{cer_large}
\end{figure}

We then evaluate the cross-entropy between the predictions and the labels and compare it to the Character Error Rate (CER) difference between our proposed method and the baseline, with the results for Speech at -10dB, for the Large configuration, presented in Fig. \ref{cer_large}. For clarity, we exclude utterances where both models have the same CER, i.e. their difference is zero.

Figure \ref{cer_large} demonstrates that VisG AV-HuBERT achieves greater improvements over baseline for sentences with higher complexity (CE > 3). This suggests that viseme-level supervision provides stronger benefits for complex utterances by guiding the model toward discriminative visual articulatory features that help resolve ambiguities in challenging contexts. Similar patterns are observed across other noise types and SNR levels, and the results on the LRS2 dataset further validate these findings. Full details are provided in the Supplementary Material.

\section{Ablation - Viseme Loss Weight}\label{sec:ablation}
In this section, we investigate the impact of the auxiliary loss weight $\alpha$. %The ablation is performed on the LRS3 test set. 
We refer readers to the Supplementary Material for ablation studies regarding noise-related hyper-parameters and the viseme sub-network architecture. % (Section \ref{sec:noise_params})  (Section \ref{sec:network})
% \subsection{Viseme Loss Weight} \label{sec:viseme_weights}
We conducted a grid search over the auxiliary loss weight $\alpha$, experimenting with both manual tuning and learnable parameters. Table \ref{tab:viseme_weight} presents WER results across the tested $\alpha$ values, averaged across noise types at each SNR level, showing the optimal configuration for Base and Large model variants. When selecting the optimal auxiliary loss weight $\alpha$, we prioritize balanced improvements across both clean and noisy conditions, especially under heavy noise, rather than optimizing exclusively for clean speech performance. This ensures VisG AV-HuBERT maintains practical robustness across diverse acoustic scenarios. 

\begin{table}[htbp]
\centering
\caption{WER across auxiliary loss weight $\alpha$. %Results average performance across noise types at each SNR level, with 
Optimal weights are indicated in bold.}
\label{tab:viseme_weight}
\scalebox{0.95}{
\begin{tabular}{|c|ccccc|c|}
\hline
\multicolumn{7}{|c|}{\textbf{Base Model}} \\ 
\hline
\textbf{Weight ($\alpha$)} & \textbf{-10 dB} & \textbf{-5 dB} & \textbf{0 dB} & \textbf{5 dB} & \textbf{10 dB} & \textbf{Clean}\\
\hline
Baseline ($\alpha$=0) & 22.32 & 13.17 & 7.66 & 5.36 & 4.51 & 4.1 \\
\hline
$\alpha$ = 0.15 & 22.93 & 13.6 & 7.78 & 5.20 & 4.42 & 3.62 \\
\textbf{$\alpha$ = 0.2} & \textbf{21.85} & \textbf{12.4} & \textbf{7.12} & \textbf{5.27} & \textbf{4.47} & \textbf{3.77}\\
$\alpha$ = 0.25 & 23.5 & 13.6 & 7.72 & 5.47 & 4.64 & 3.78 \\
$\alpha$ = 0.3 & 23.16 & 13.4 & 7.63 & 5.3 & 4.47 & 3.74 \\
$\alpha$ = 0.35 & 23.31 & 13.14 & 7.44 & 5.43 & 4.6 & 3.76 \\
learnable & 23.75 & 13.55 & 7.43 & 5.29 & 4.29 & 3.63 \\
\hline
\multicolumn{7}{|c|}{\textbf{Large Model}} \\ 
\hline
Baseline ($\alpha$=0) & 15.91 & 6.965 & 3.03 & 2.07 & 1.72 & 1.4 \\
\hline
% $\alpha$ = 0.1 & 14.20 & 9.15 & 6.68 & 5.36 & 4.70 \\
\textbf{$\alpha$ = 0.15} & \textbf{13.96} & \textbf{7.89} & \textbf{3.31} & \textbf{2.14} & \textbf{1.67} & \textbf{1.38} \\
$\alpha$ = 0.25 & 15.87 & 7.58 & 3.49 & 2.19 & 1.78 & 1.54 \\
$\alpha$ = 0.3 & 15.55 & 7.49 & 3.59 & 2.19 & 1.79 & 1.49 \\
\hline
\end{tabular}
}
\end{table}

We find that the optimal values for the Base and the Large models differ, with the viseme weight for the Base model being equal to 0.2, while for the Large model equal to 0.15. We observed that learnable weights, i.e. dynamic parameters adjusted during updates, improved performance on clean speech but offered no advantage under noisy conditions. Our experiments demonstrate a clear optimal value for the auxiliary loss weight $\alpha$. Excessively high weights cause the secondary loss to act as noise rather than regularization, degrading the model's ability to learn useful representations.

\section{Conclusion and Future Work}\label{sec:conclusion}
% Bullet - Point List of content.
% \begin{itemize}
%     \item Method may not improve by a margin, but sheds light to a possible avenue for improvement, especially under noisy conditions.
%     \item Future work could include different weighting schema, multi-lingual setting or low-resourced languages, introduction of more tasks
%     \item Limitations include viseme mapping, optimal weight, performance in babble
% \end{itemize}

We have proposed VisG AV-HuBERT, a viseme-guided multi-task fine-tuning framework based on AV-HuBERT. VisG AV-HuBERT explicitly leverages visual articulatory information, to enhance AV-HuBERT's encoder representations.
%and to better utilize the visual modality
Our method achieves improvements in noisy conditions, while being able to maintain performance in clean conditions, demonstrating that explicit viseme modelling enhances robustness in noise. 

% While our method achieves modest but consistent improvements over baseline in clean conditions, it demonstrates that explicit viseme modelling enhances robustness in noisy conditions. 

Our experiments on LRS3 demonstrate consistent improvements over baseline AV-HuBERT, particularly under challenging acoustic conditions, i.e. in -10dB, -5dB. At severe noise levels (-10 dB SNR), VisG AV-HuBERT Large achieves a 51.4\% relative WER reduction for Speech noise (from 13.59\% to 6.60\%). Error analysis reveals substantial reductions in substitution errors across all noise types, demonstrating improved phonetic discrimination. Cross-dataset evaluation on LRS2 confirms generalization capability, with the Large model achieving 9.925\% WER compared to baseline 10.3\%, in clean conditions, while in noise, similar patterns and improvements are noted as well.

Our results show that explicit viseme modelling enhances encoder representations by guiding models toward discriminative visual articulatory features. In contrast to current trends focusing on increasingly sophisticated LLM decoders, our work demonstrates that encoder improvements through multi-task auxiliary objectives provide a complementary path to advancing AVSR performance. This approach is particularly valuable for noise-robust speech recognition, where visual information becomes critical for disambiguating phonetically similar sounds.

While our method shows promising results, several limitations warrant investigation. First, the optimal $\alpha$ varies across model sizes, suggesting need for adaptive weighting schemes. Second, performance on Babble noise remains challenging, indicating opportunities for noise-type-specific modelling. Finally, further investigation is needed to understand how the phoneme-to-viseme mapping interacts with and influences overall model performance.

Future work could explore extending this framework to multilingual settings with language-specific viseme mappings. We believe that low-resourced languages might benefit from our approach, as it extends the capacity of the data. Further investigation into dynamic loss weight strategies, may also benefit our framework. Finally, changing or including more auxiliary tasks, such as phoneme prediction or articulatory feature classification could be used to strengthen encoder representations further.

\subsubsection{Acknowledgement} This publication has emanated from research conducted with the financial support of Taighde Éireann – Research Ireland under Grant number 22/FFP-A/11059.


\begin{thebibliography}{99}

\bibitem{llama_avsr}
Cappellazzo, U., Kim, M., Chen, H., Ma, P., Petridis, S., Falavigna, D., Brutti, A., Pantic, M.: Large language models are strong audio-visual speech recognition learners. In: ICASSP 2025, pp. 1--5 (2025).

% \bibitem{matryoshka}
% Cappellazzo, U., Kim, M., Petridis, S.: Adaptive audio-visual speech recognition via Matryoshka-based multimodal LLMs. In: ASRU 2025 (2025).

\bibitem{sun2024video-salmonn} %, Yu, W., Tang, C., Chen, X., Tan, T., Li, W., Lu, L., Ma, Z., Wang, Y., Zhang, C.
Sun, G. et al.: video-SALMONN: Speech-enhanced audio-visual large language models. In: Proceedings of ICML 2024, pp. 1--20. JMLR.org (2024).

\bibitem{cappellazzo25_interspeech}
Cappellazzo, U., Kim, M., Petridis, S., Falavigna, D., Brutti, A.: Scaling and enhancing LLM-based AVSR: A sparse mixture of projectors approach. In: Interspeech 2025, pp. 1823--1827 (2025).

\bibitem{zeroshotavsr}
Yeo, J.H., Kim, M., Kim, C.W., Petridis, S., Ro, Y.M.: Zero-AVSR: Zero-shot audio-visual speech recognition with LLMs by learning language-agnostic speech representations. In: ICCV 2025.

\bibitem{mms_llama}
Yeo, J.H., Rha, H., Park, S.J., Ro, Y.M.: MMS-LLaMA: Efficient LLM-based audio-visual speech recognition with minimal multimodal speech tokens. In: Findings of the Association for Computational Linguistics: ACL 2025 (2025).

\bibitem{vspllm}
Yeo, J., Han, S., Kim, M., Ro, Y.M.: Where visual speech meets language: VSP-LLM framework for efficient and context-aware visual speech processing. In EMNLP 2024, pp. 11391--11406 %(2024). % Findings of the Association for Computational Linguistics:

\bibitem{vallr}
Thomas, M., Fish, E., Bowden, R.: VALLR: Visual ASR language model for lip reading. In: ICCV 2025.

\bibitem{avhubert}
Shi, B., Hsu, W.-N., Lakhotia, K., Mohamed, A.: Learning audio-visual speech representation by masked multimodal cluster prediction. In: ICLR 2022.

\bibitem{auto_avsr}
Ma, P., Haliassos, A., Fernandez-Lopez, A., Chen, H., Petridis, S., Pantic, M.: Auto-AVSR: Audio-visual speech recognition with automatic labels. In: ICASSP 2023, pp. 1--5 %(2023).

\bibitem{whisper}
Radford, A., Kim, J.W., Xu, T., Brockman, G., Mcleavey, C., Sutskever, I.: Robust speech recognition via large-scale weak supervision. In: ICML 2023, vol. 202, pp. 28492--28518. % PMLR (2023)

\bibitem{wav2vec2}
Baevski, A., Zhou, H., Mohamed, A., Auli, M.: wav2vec 2.0: A framework for self-supervised learning of speech representations. In: NeurIPS 2020.

\bibitem{visemes}
Fisher, C.G.: Confusions among visually perceived consonants. Journal of Speech and Hearing Research \textbf{11}(4), 796--804 (1968).

\bibitem{ma_probing_2021}
Ma, D., Ryant, N., Liberman, M.: Probing acoustic representations for phonetic properties. In: ICASSP 2021, pp. 311--315.

\bibitem{pasad_layerwise_2021}
Pasad, A., Chou, J.-C., Livescu, K.: Layer-wise analysis of a self-supervised speech representation model. In: ASRU 2021, pp. 914--921.

% \bibitem{cormac_english_domaininformed_2022}
% English, P.C., Kelleher, J.D., Carson-Berndsen, J.: Domain-informed probing of wav2vec 2.0 embeddings for phonetic features. In: SIGMORPHON 2022, pp. 83--91.

\bibitem{english23interspeech}
English, P.C., Kelleher, J.D., Carson-Berndsen, J.:
Discovering phonetic feature event patterns in transformer embeddings.
In: Interspeech 2023, pp. 4733--4737. %(2023).
% \doi{10.21437/Interspeech.2023-1985}


\bibitem{seyssel_2022}
de Seyssel, M., Lavechin, M., Adi, Y., Dupoux, E., Wisniewski, G.: Probing phoneme, language and speaker information in unsupervised speech representations. In: Interspeech 2022, pp. 1402--1406.

\bibitem{pasad_comparative_2023}
Pasad, A., Shi, B., Livescu, K.: Comparative layer-wise analysis of self-supervised speech models. In: ICASSP 2023, pp. 1--5.

\bibitem{papadopoulos2025visemes}
Papadopoulos, A., Harte, N.: Interpreting the role of visemes in audio-visual speech recognition. In: ASRU 2025.

\bibitem{caruana1997mtl}
Caruana, R.: Multi-task learning. Machine Learning \textbf{28}, 41--75 (1997).

\bibitem{univpm}
Hu, Y., Li, R., Chen, C., Qin, C., Zhu, Q.-S., Chng, E.S.: Hearing lips in noise: Universal viseme-phoneme mapping and transfer for robust audio-visual speech recognition. In: ACL 2023, pp. 15213--15232.

\bibitem{grattafiori2024llama3}
Grattafiori, A., et al.: The Llama 3 Herd of Models. arXiv preprint arXiv:2407.21783 (2024).

\bibitem{lora}
Hu, E.J., et al.: LoRA: Low-rank adaptation of large language models. In: ICLR 2022.

\bibitem{cappelletta2012phonemetoviseme}
Cappelletta, L., Harte, N.: Phoneme-to-viseme mapping for visual speech recognition. In: ICPRAM 2012.

\bibitem{guan2025mllm} %  
Guan, Y., Trinh, V.A., Voleti, V., Whitehill, J.: MLLM-based speech recognition: When and how is multimodality beneficial? arXiv preprint arXiv:2507.19037 (2025).

\bibitem{lrs3}
Afouras, T., Chung, J.S., Zisserman, A.: LRS3-TED: A large-scale dataset for visual speech recognition. arXiv:1809.00496 (2018).

\bibitem{noisy_avhubert}
Shi, B., Hsu, W.-N., Mohamed, A.: Robust self-supervised audio-visual speech recognition. In: Interspeech 2022, pp. 2118--2122.

\bibitem{MUSAN}
Snyder, D., Chen, G., Povey, D.: MUSAN: A music, speech, and noise corpus. arXiv:1510.08484 (2015).

\bibitem{mcauliffe17_interspeech}
McAuliffe, M., Socolof, M., Mihuc, S., Wagner, M., Sonderegger, M.: Montreal forced aligner: Trainable text-speech alignment using Kaldi. In: Interspeech 2017, pp. 498--502.

% \bibitem{gorman2011prosodylab}
% Gorman, K., Howell, J., Wagner, M.: Prosodylab-aligner: A tool for forced alignment of speech. Canadian Acoustics \textbf{39}(3), 192--193 (2011).

\bibitem{LRS2}
Afouras, T., Chung, J.S., Senior, A., Vinyals, O., Zisserman, A.: Deep audio-visual speech recognition. arXiv:1809.02108 (2018).

\bibitem{hsu2022uhubert}
Hsu, W.-N., Shi, B.: u-HuBERT: Unified mixed-modal speech pretraining and zero-shot transfer to unlabeled modality. In: NeurIPS 2022, pp. 21157--21170.

\bibitem{haliassos2025usr}
Haliassos, A., Mira, R., Chen, H., Landgraf, Z., Petridis, S., Pantic, M.: Unified speech recognition: A single model for auditory, visual, and audiovisual inputs. In: NeurIPS 2024, pp. 1--27.

\bibitem{multiavsr}
Torrie, S., Wright, K., Lee, D.-J.: MultiAVSR: Robust speech recognition via supervised multi-task audio–visual learning. Electronics \textbf{14}(12), 2310 (2025).

\bibitem{jeffers}
Jeffers, J., Barley, M.: Speechreading (Lipreading). Thomas (1971).

\bibitem{bear2017decodingvisemesimprovingmachine}
Bear, H.L., Harvey, R.: Decoding visemes: Improving machine lip-reading. In: ICASSP 2016, pp. 2009--2013.

% \bibitem{bozkurt}
% Bozkurt, E., Erdem, C., Erzin, E., Erdem, T., Özkan, M.: Comparison of phoneme and viseme based acoustic units for speech driven realistic lip animation. In: SPC 2007, pp. 1--4.

\bibitem{hazen}
Hazen, T.J.: Visual model structures and synchrony constraints for audio-visual speech recognition. IEEE Transactions on Audio, Speech, and Language Processing \textbf{14}(3), 1082--1089 (2006).

\bibitem{leemap}
Lee, S., Yook, D.: Audio-to-visual conversion using hidden Markov models. In: PRICAI 2002, pp. 563--570.

\bibitem{bahdanau2016attention}
Bahdanau, D., Chorowski, J., Serdyuk, D., Brakel, P., Bengio, Y.: End-to-end attention-based speech recognition. In: ICASSP 2016, pp. 4945--4949.

\bibitem{ctc_loss}
Graves, A., Fernández, S., Gomez, F., Schmidhuber, J.: Connectionist temporal classification: Labelling unsegmented sequence data with RNNs. In: ICML 2006, pp. 369--376.

\bibitem{hybrid_ctc_ce} %
Watanabe, S., Hori, T., Kim, S., Hershey, J.R., Hayashi, T.: Hybrid CTC/Attention architecture for end-to-end speech recognition. IEEE JSTSP \textbf{11}(8), 1240--1253 (2017).

\bibitem{stewart2014robust}
Stewart, D., Seymour, R., Pass, A., Ming, J.: Robust audio-visual speech recognition under noisy audio-video conditions. IEEE Transactions on Cybernetics \textbf{44}(2), 175--184 (2014).

\bibitem{lim2025improving}
Lim, D., Kim, Y., Kim, D.-H., Yang, D.-H., Chang, J.-H.: Improving noise-robust AVSR via router-gated cross-modal feature fusion. In: ASRU 2025.

\bibitem{lin2025uncovering}
Lin, Z., Harte, N.: Uncovering the visual contribution in audio-visual speech recognition. In: ICASSP 2025, pp. 1--5.

\bibitem{sterpu2020avalign}
Sterpu, G., Saam, C., Harte, N.: How to teach DNNs to pay attention to the visual modality in speech recognition. IEEE/ACM TASLP \textbf{28}, 1052--1064 (2020).%\url{https://doi.org/10.1109/TASLP.2020.2980436}


\end{thebibliography}
%

\section{Supplementary Material}

% Include WER tables with the different experiments?\
% Examples of where the model improves or decreases as well?
% More Perplexity Sentences ?

\subsection{Detailed WER Results}
In Table \ref{tab:lrs2_noise} we present the WER in noisy conditions for the LRS2 dataset, ranging from -10dB to 10dB, for Babble, Speech, Music and Random Noise types.

\begin{table}[htbp]
\centering
\caption{WER Comparison under Different Noise Types for LRS2}
\label{tab:lrs2_noise}
% \resizebox{0.9\textwidth}{!}{
\begin{tabular}{|l|c|ccccc|}
\hline
\textbf{Model} & \textbf{Noise Type} & \textbf{-10 dB} & \textbf{-5 dB} & \textbf{0 dB} & \textbf{5 dB} & \textbf{10 dB} \\
\hline
AV-HuBERT Base & \multirow{4}{*}{\centering Babble} & 52.72 & 37.73 & 23.78 & 17.8 & 14.895 \\
VisG AV-HuBERT Base &  & 53.33 & \textbf{36.78} & \textbf{23.09} & \textbf{17.61} & 15.42 \\
\cline{1-1}\cline{3-7}
AV-HuBERT Large &  & 43.06 & 30.18 & 18.59 & 15.57 & 14.38 \\
VisG AV-HuBERT Large &  & \textbf{41.65} & \textbf{28.76$^{**}$} & 19.17 & \textbf{15.06} & \textbf{13.81} \\
\hline
AV-HuBERT Base & \multirow{4}{*}{\centering Speech} & 39.89 & 26.24 & 19.71 & 16.14 & 14.79 \\
VisG AV-HuBERT Base &  & \textbf{31.09$^{**}$} & \textbf{23.74$^{**}$} & \textbf{19.07} & 16.71 & 15.25 \\
\cline{1-1}\cline{3-7}
AV-HuBERT Large &  & 27.1 & 19.13 & 15.48 & 14.25 & 13.74 \\
VisG AV-HuBERT Large &  & \textbf{24.13$^{**}$} & \textbf{17.67$^{**}$} & 15.61 & \textbf{13.65} & \textbf{13.42} \\
\hline
AV-HuBERT Base & \multirow{4}{*}{\centering Music} & 30.5 & 22.26 & 17.69 & 15.26 & 14.11 \\
VisG AV-HuBERT Base &  & \textbf{30.07} & \textbf{22.1} & \textbf{17.34} & 15.3 & 14.32 \\
\cline{1-1}\cline{3-7}
AV-HuBERT Large &  & 23.6 & 17.58 & 15.07 & 14.14 & 14.02 \\
VisG AV-HuBERT Large &  & \textbf{22.76} & \textbf{17.34} & \textbf{14.53} & \textbf{13.72} & \textbf{13.24} \\
\hline
AV-HuBERT Base & \multirow{4}{*}{\centering Noise} & 28.21 & 21.45 & 17.14 & 15.25 & 14.31 \\
VisG AV-HuBERT Base &  & 29.38 & \textbf{21.36} & 17.21 & 15.46 & 14.32 \\
\cline{1-1}\cline{3-7}
AV-HuBERT Large &  & 23.42 & 17.67 & 14.88 & 14.5 & 14.43 \\
VisG AV-HuBERT Large &  & \textbf{22.21} & \textbf{17.1} & 14.91 & \textbf{13.65} & \textbf{13.33} \\
\hline
\end{tabular}
% }
\end{table}

Interestingly, the performance of VisG AV-HuBERT in LRS2 is more nuanced. VisG AV-HuBERT Large maintains consistent improvements across most conditions, particularly for Speech noise (22\% relative improvement at -10 dB) and Music noise. However, the base model shows mixed results: while achieving notable gains for Speech noise (22\% relative improvement at -10 dB), performance degrades slightly for Babble and Noise categories at extreme SNR levels. This suggests that the Base model's improvements are partially dataset-specific, whereas the Large model demonstrates more robust cross-dataset generalization. The stronger Speech noise improvements on both datasets confirm that viseme guidance particularly benefits speech-on-speech discrimination, where visual articulatory cues provide consistent disambiguating information regardless of dataset domain. The LRS2 results validate that explicit viseme modelling enhances encoder representations, though the extent of improvement depends on model capacity and acoustic conditions.

\subsubsection{Results under same noise hyper-parameter settings}

We decided to fine-tune the original AV-HuBERT architecture with our noise hyper-parameters and provide the results in \ref{tab:noise_comparison}.

\begin{table}[htbp]
    \centering
    \caption{WER comparison in LRS3 between the proposed model and AV-HuBERT, trained under same noise hyper-parameter settings.}
    \label{tab:noise_comparison}
    % \resizebox{\textwidth}{!}{
    \begin{tabular}{|l|c|ccccc|}
    \hline
    \textbf{Model} & \textbf{Noise Type} & \textbf{-10 dB} & \textbf{-5 dB} & \textbf{0 dB} & \textbf{5 dB} & \textbf{10 dB} \\
    \hline
    AV-HuBERT Base & \multirow{4}{*}{\centering Babble} & 41.70 & 22.86 & 10.81 & 6.60 & 5.05 \\
    VisG AV-HuBERT Base &  & 42.55 & \textbf{22.60} & \textbf{10.18} & \textbf{6.03} & \textbf{4.65} \\
    \cline{1-1}\cline{3-7}
    AV-HuBERT Large &  & 33.26 & 14.76 & 4.60 & 2.37 & 1.62 \\
    VisG AV-HuBERT Large &  & \textbf{31.03} & \textbf{14.05} & 4.92 & 2.51 & 1.87 \\
    \hline
    AV-HuBERT Base & \multirow{4}{*}{\centering Speech} & 13.05 & 8.50 & 6.55 & 5.54 & 4.93 \\
    VisG AV-HuBERT Base &  & 13.55 & 8.78 & 6.60 & \textbf{5.33} & \textbf{4.66} \\
    \cline{1-1}\cline{3-7}
    AV-HuBERT Large &  & 6.29 & 3.49 & 2.52 & 2.06 & 1.62 \\
    VisG AV-HuBERT Large &  & 6.60 & 3.83 & \textbf{2.50} & \textbf{1.94} & \textbf{1.61} \\
    \hline
    AV-HuBERT Base & \multirow{4}{*}{\centering Music} & 15.41 & 9.42 & 6.06 & 5.09 & 4.49 \\
    VisG AV-HuBERT Base &  & 16.60 & \textbf{9.36} & 6.12 & \textbf{5.00} & \textbf{4.30} \\
    \cline{1-1}\cline{3-7}
    AV-HuBERT Large &  & 9.18 & 4.30 & 2.58 & 1.84 & 1.98 \\
    VisG AV-HuBERT Large &  & 9.62 & 4.63 & \textbf{2.52} & 1.92 & \textbf{1.61} \\
    \hline
    AV-HuBERT Base & \multirow{4}{*}{\centering Noise} & 15.24 & 8.95 & 6.20 & 4.86 & 4.50 \\
    VisG AV-HuBERT Base &  & \textbf{14.72} & \textbf{8.87} & \textbf{5.59} & 5.27 & 4.66 \\
    \cline{1-1}\cline{3-7}
    AV-HuBERT Large &  & 9.69 & 5.61 & 2.68 & 1.83 & 2.07 \\
    VisG AV-HuBERT Large &  & \textbf{8.59} & \textbf{4.18} & \textbf{2.29} & 1.94 & \textbf{1.68} \\
    \hline
    \end{tabular} 
    % }
\end{table}

Table \ref{tab:noise_comparison} presents an ablation study comparing VisG AV-HuBERT against baseline AV-HuBERT when both are trained with identical noise augmentation parameters (p=0.3, SNR=-2.5 dB). The results reveal that the viseme auxiliary task provides selective rather than universal benefits. For the Base model, improvements are observed primarily for Babble noise at moderate SNR levels (-5 dB to 10 dB) and for the Noise category across most conditions. The Large model shows similar patterns, with consistent improvements for the Noise category but mixed results for Speech and Music noise. Notably, both models show performance degradation at extreme SNR levels (-10 dB) for several noise types, suggesting that the auxiliary viseme task is most effective when acoustic signals retain sufficient information for the model to leverage visual articulatory cues. These results confirm that our overall improvements stem from the combination of optimized training hyperparameters and targeted viseme guidance, rather than the auxiliary task alone.

\subsection{Ablation: Network-Depth} \label{sec:network}
We experimented with 1, 2 and 3 sequences of Linear Layer, Layer Normalization, GELU activation and Dropout, with p=0.3. In Table \ref{tab:layer_ablation} we present the different configurations we tested under for the Base and the Large model.

\begin{table}[htbp]
    \centering
    \caption{WER across different viseme sub-network configurations. Results average performance across noise types at each SNR level. Optimal configuration highlighted in bold.}
    \label{tab:layer_ablation}
    \begin{tabular}{|c|c|ccccc|c|}
    \hline
    \multicolumn{8}{|c|}{\textbf{Base Model}} \\ 
    \hline
    \textbf{Number of Layers} & \textbf{Weight ($\alpha$)} & \textbf{-10 dB} & \textbf{-5 dB} & \textbf{0 dB} & \textbf{5 dB} & \textbf{10 dB} & \textbf{Clean}\\
    \hline
    1 & $\alpha$ = 0.2 & 24.11 & 13.78 & 7.58 & 5.4398 & 4.57 & 3.75 \\
    \hline
    \textbf{2} & \textbf{$\alpha$ = 0.2} & \textbf{21.85} & \textbf{12.4} & \textbf{7.12} & \textbf{5.27} & \textbf{4.47} & \textbf{3.77}\\
    \hline
    \multicolumn{8}{|c|}{\textbf{Large Model}} \\ 
    \hline
    2 & $\alpha$ = 0.15 & 15.48 & 7.2 & 3.4 & 2.06 & 1.63 & 1.27 \\
    \hline
    \textbf{3} & \textbf{$\alpha$ = 0.15} & \textbf{13.96} & 1\textbf{7.89} & \textbf{3.31} & \textbf{2.14} & \textbf{1.67} & \textbf{1.38} \\
    \hline
    \end{tabular}
\end{table}

The Base model achieves optimal performance with 2 layers, while the Large model requires 3 layers. This difference in optimal depth, combined with the different auxiliary loss weights ($\alpha$=0.2 for Base, $\alpha$=0.15 for Large), reflects the varying representational capacities of the two architectures. The Larger model benefits from a deeper viseme sub-network that can learn more complex mappings between encoder representations and viseme classes, while the Base model performs best with a shallower network that provides sufficient guidance without introducing excessive parameters.

\subsection{Ablation: Noise hyper-parameters} \label{sec:noise_params}
It has become standard when fine-tuning an AVSR model to augment the audio files, by randomly mixing noise at 0 dB with a probability of 0.25 \cite{avhubert}. We evaluated multiple SNR and dB levels and the results are shown in Table \ref{tab:noise_table}.

\begin{table}[htbp]
\centering
\caption{WER across different noise hyper-parameter combinations.Results average performance across noise types at each SNR level. Optimal configuration highlighted in bold.}
\label{tab:noise_table}
\begin{tabular}{|c|c|ccccc|c|}
\hline
\multicolumn{8}{|c|}{\textbf{Base Model}} \\ 
\hline
\textbf{Noise Probability} & \textbf{SNR Level (dB)} & \textbf{-10 dB} & \textbf{-5 dB} & \textbf{0 dB} & \textbf{5 dB} & \textbf{10 dB} & \textbf{Clean}\\
\hline
Baseline ($p$=0.25) & 0  & 22.32 & 13.17 & 7.66 & 5.36 & 4.51 & 4.1 \\
\hline
$p$=0.25 & -2.5  & 21.55 & 12.8 & 7.41 & 5.5 & 4.66 & 3.92 \\
\textbf{$p$=0.3} & \textbf{-2.5} & \textbf{21.85} & \textbf{12.4} & \textbf{7.12} & \textbf{5.27} & \textbf{4.47} & \textbf{3.77}\\
$p$=0.5 & 0  & 23.19 & 12.92 & 7.11 & 5.4 & 4.5 & 3.8 \\
\hline
\multicolumn{8}{|c|}{\textbf{Large Model}} \\ 
\hline
Baseline ($p$=0.25) & 0 & 15.91 & 6.965 & 3.03 & 2.07 & 1.72 & 1.4 \\
\hline
$p$=0.25 & -2.5  & 14.97 & 7.14 & 3.32 & 2.27 & 1.88 & 1.485 \\
\textbf{$p$=0.3} & \textbf{-2.5} & \textbf{13.96} & \textbf{7.89} & \textbf{3.31} & \textbf{2.14} & \textbf{1.67} & \textbf{1.38} \\
$p$=0.4 & 0  & 14.72 & 7.13 & 3.16 & 2.1 & 1.69 & 1.4 \\
\hline
\end{tabular}
\end{table}

Our experiments demonstrate that noise augmentation hyperparameters, mixing probability p and SNR level, significantly impact WER performance. The optimal configuration for both Base and Large models uses p=0.3 and SNR=-2.5 dB, improving over the baseline across all noise conditions. However, excessive augmentation (e.g., p=0.5 or very low SNR) provides diminishing returns and can destabilize training, as evidenced by the p=0.5 configuration showing minimal if any improvements.

\subsection{Character Error Rate Results}
In Table \ref{tab:cer_noise}, we present the CER results under noise, for the LRS3 dataset, ranging from -10dB to 10dB, for Babble, Speech, Music and Random Noise.

\begin{table}[htbp]
\centering
\caption{CER (\%) Comparison under Different Noise Types for LRS3}
\label{tab:cer_noise}
\scalebox{0.9}{
\begin{tabular}{|l|c|ccccc|}
\hline
\textbf{Model} & \textbf{Noise Type} & \textbf{-10 dB} & \textbf{-5 dB} & \textbf{0 dB} & \textbf{5 dB} & \textbf{10 dB} \\
\hline
AV-HuBERT Base & \multirow{4}{*}{\centering Babble} & 31.5 & 17.25 & 7.09 & 3.27 & 2.5 \\
VisG AV-HuBERT Base &  & 32.43 & \textbf{16.15$^{*}$} & \textbf{6.66$^{*}$} & 3.50 & \textbf{2.49$^{*}$} \\
\cline{1-1}\cline{3-7}
AV-HuBERT Large &  & 23.47 & 10.57 & 3.60 & 1.70 & 1.21 \\
VisG AV-HuBERT Large &  & 24.25 & 11.10 & 3.66 & 1.74 & 1.24 \\
\hline

AV-HuBERT Base & \multirow{4}{*}{\centering Speech} & 10.39 & 6.24 & 3.96 & 2.76 & 2.48 \\
VisG AV-HuBERT Base &  & \textbf{8.92$^{**}$} & \textbf{5.76$^{*}$} & 4.05 & 3.06 & 2.60 \\
\cline{1-1}\cline{3-7}
AV-HuBERT Large &  & 12.14 & 3.64 & 2.51 & 1.38 & 1.27 \\
VisG AV-HuBERT Large &  & \textbf{5.10$^{**}$} & \textbf{2.62$^{**}$} & \textbf{1.60$^{**}$} & \textbf{1.19} & \textbf{1.02} \\
\hline

AV-HuBERT Base & \multirow{4}{*}{\centering Music} & 12.22 & 6.33 & 3.66 & 2.62 & 2.26 \\
VisG AV-HuBERT Base &  & \textbf{11.49} & \textbf{5.86} & 3.67 & 2.64 & \textbf{2.24} \\
\cline{1-1}\cline{3-7}
AV-HuBERT Large &  & 7.76 & 3.61 & 1.80 & 1.31 & 1.10 \\
VisG AV-HuBERT Large &  & \textbf{7.11} & \textbf{3.35} & \textbf{1.62} & \textbf{1.31} & \textbf{0.98} \\
\hline

AV-HuBERT Base & \multirow{4}{*}{\centering Noise} & 10.65 & 5.42 & 3.51 & 2.78 & 2.20 \\
VisG AV-HuBERT Base &  & \textbf{10.32} & 5.85 & \textbf{3.14$^{*}$} & \textbf{2.60$^{*}$} & 2.32 \\
\cline{1-1}\cline{3-7}
AV-HuBERT Large &  & 7.54 & 3.49 & 1.75 & 1.16 & 0.98 \\
VisG AV-HuBERT Large &  & \textbf{6.49} & \textbf{3.12} & \textbf{1.61} & 1.20 & \textbf{0.89} \\
\hline
\end{tabular}
}
\end{table}

CER results in Table \ref{tab:cer_noise} corroborate the WER findings, demonstrating consistent improvements with viseme guidance across character-level predictions. VisG AV-HuBERT Large achieves the most substantial CER reductions under Speech noise, with a 58.0\% relative improvement at -10 dB SNR (12.14\% to 5.10\%, p < 0.01), 28.0\% at -5 dB (3.64\% to 2.62\%, p<0.01), and 36.3\% at 0 dB (2.51\% to 1.60\%, p<0.01). These character-level improvements are proportionally larger than the corresponding WER reductions, suggesting that viseme guidance particularly helps prevent character substitution and insertion errors that compound into word-level mistakes. Consistent with WER trends, Music and Random Noise conditions show moderate but reliable CER improvements, while Babble noise at extreme SNR levels remains challenging for the Large model. The alignment between WER and CER improvements across noise types and SNR levels validates that viseme-guided learning enhances recognition robustness at both character and word granularities.

Table \ref{tab:cer_noise_lrs2} presents the CER results, for the various noise conditions in LRS2.

\begin{table}[htbp]
\centering
\caption{CER (\%) Comparison under Different Noise Types for LRS2}
\label{tab:cer_noise_lrs2}
\scalebox{0.9}{
\begin{tabular}{|l|c|ccccc|}
\hline
\textbf{Model} & \textbf{Noise Type} & \textbf{-10 dB} & \textbf{-5 dB} & \textbf{0 dB} & \textbf{5 dB} & \textbf{10 dB} \\
\hline
AV-HuBERT Base & \multirow{4}{*}{\centering Babble} & 37.92 & 26.04 & 15.74 & 11.59 & 9.62 \\
VisG AV-HuBERT Base &  & 38.46 & \textbf{26.01} & \textbf{15.66} & 11.84 & 10.25 \\
\cline{1-1}\cline{3-7}
AV-HuBERT Large &  & 32.04 & 21.86 & 13.72 & 11.32 & 10.47 \\
VisG AV-HuBERT Large &  & \textbf{30.66} & \textbf{20.73} & 14.03 & \textbf{10.79} & \textbf{10.10} \\
\hline

AV-HuBERT Base & \multirow{4}{*}{\centering Speech} & 28.71 & 18.09 & 13.25 & 10.62 & 9.75 \\
VisG AV-HuBERT Base &  & \textbf{21.92$^{**}$} & \textbf{16.21$^{*}$} & \textbf{12.74} & 11.17 & 10.26 \\
\cline{1-1}\cline{3-7}
AV-HuBERT Large &  & 20.51 & 14.33 & 11.31 & 10.48 & 10.11 \\
VisG AV-HuBERT Large &  & \textbf{17.79$^{**}$} & \textbf{12.97$^{**}$} & \textbf{11.26} & \textbf{10.00} & \textbf{9.88} \\
\hline

AV-HuBERT Base & \multirow{4}{*}{\centering Music} & 21.25 & 14.36 & 11.55 & 9.94 & 9.24 \\
VisG AV-HuBERT Base &  & \textbf{20.43} & 14.97 & 11.64 & 10.31 & 9.76 \\
\cline{1-1}\cline{3-7}
AV-HuBERT Large &  & 17.40 & 12.90 & 11.11 & 10.38 & 10.23 \\
VisG AV-HuBERT Large &  & \textbf{16.64} & \textbf{12.38} & \textbf{10.64} & \textbf{10.06} & \textbf{9.86} \\
\hline

AV-HuBERT Base & \multirow{4}{*}{\centering Noise} & 19.52 & 14.64 & 11.28 & 10.03 & 9.46 \\
VisG AV-HuBERT Base &  & 20.34 & 14.74 & 11.43 & 10.27 & 9.47 \\
\cline{1-1}\cline{3-7}
AV-HuBERT Large &  & 17.23 & 13.05 & 10.97 & 10.72 & 10.54 \\
VisG AV-HuBERT Large &  & \textbf{16.52} & \textbf{12.40} & \textbf{10.76} & \textbf{10.15} & \textbf{9.71} \\
\hline
\end{tabular}
}
\end{table}

VisG AV-HuBERT demonstrates the most substantial CER reductions under Speech noise conditions, where the Base model achieves 23.7\% relative improvement at -10 dB SNR (28.71\% to 21.92\%, p<0.01) and the Large model shows 13.3\% relative improvement (20.51\% to 17.79\%, p<0.01). Under Babble noise, the Base model exhibits minimal improvement or slight degradation across most SNR levels, while the Large model achieves consistent reductions except at 0 dB, where CER increases slightly from 13.72\% to 14.03\%. For Music noise, the Base model shows improvements only at -10 dB (21.25\% to 20.43\%), while the Large model demonstrates consistent CER reductions across all SNR levels. Random Noise conditions show minimal improvement for the Base model, whereas the Large model consistently outperforms the baseline, particularly at extreme SNR levels. These results suggest that viseme guidance provides stronger benefits in larger model architectures when evaluated on the LRS2 dataset, with Speech noise discrimination remaining the most responsive condition to auxiliary viseme supervision. The higher absolute CER values observed across all conditions (ranging from 9-38\%) reflect the increased difficulty of the new dataset.

\subsection{Error Analysis}

\begin{figure}[htbp]
    \centering
    \includegraphics[width=0.8\textwidth]{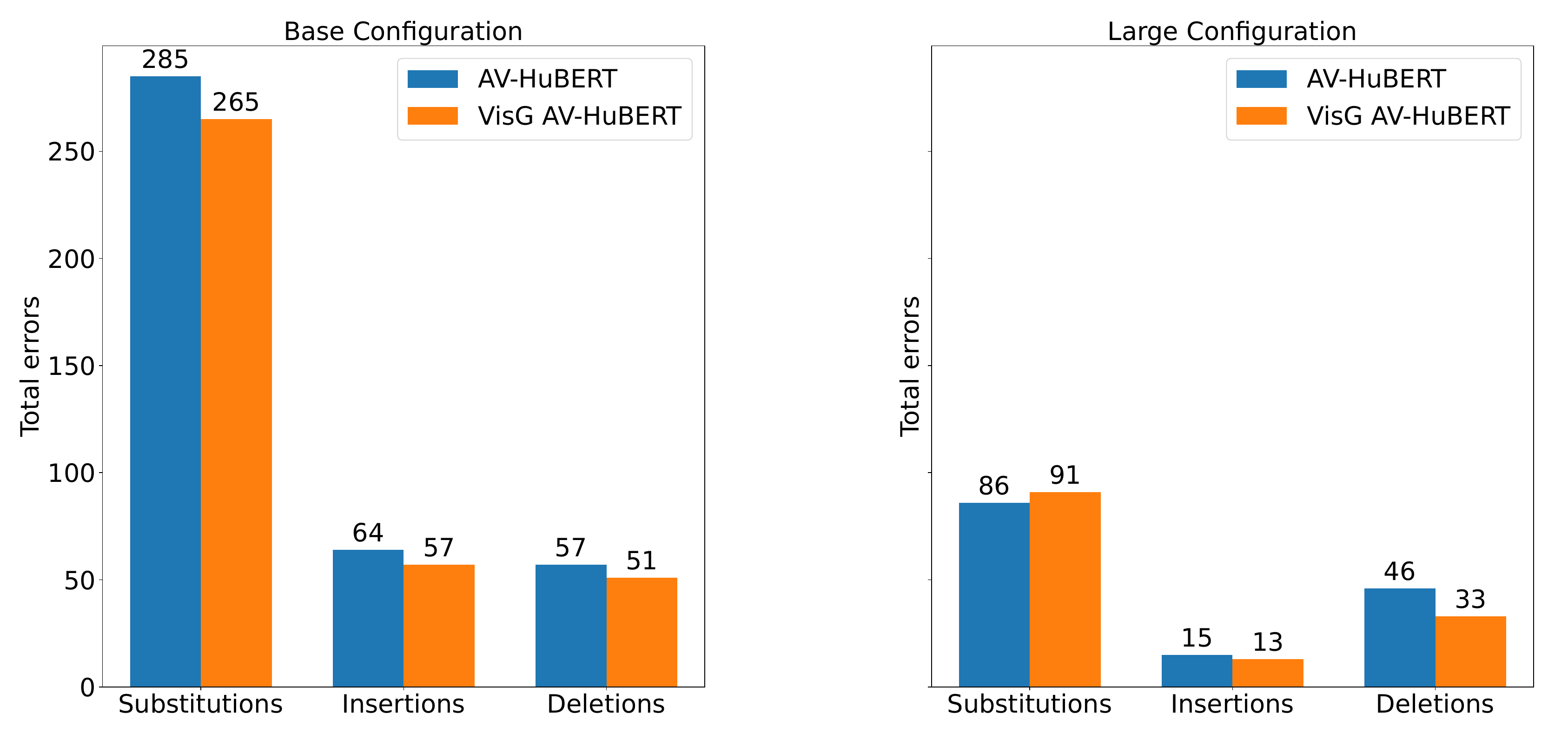}
    \caption{Errors under clean conditions for Base (left) and Large (right) models.}
    \label{fig:clean_sids}
\end{figure}

Figure \ref{fig:clean_sids} shows Substitution, Insertion, and Deletion (S/I/D) error counts under clean conditions for both Base and Large model configurations. The Base model demonstrates consistent improvements across all error categories under clean conditions, with particularly notable substitution error reductions. In contrast, the Large model exhibits a different error distribution: while insertion errors decrease substantially, substitution errors increase slightly compared to baseline. This trade-off results in modest overall WER gains (1.4\% vs. 1.38\%), suggesting that in clean acoustic conditions, the auxiliary viseme task may over-constrain the Large model's predictions, leading to hypotheses that avoid insertions but occasionally miss-classify phonetically similar sounds.

% Building on Section \ref{sec:error_types}, we provide the Error Difference breakdown for the Base model, in Figure \ref{fig:sids_base}.

\begin{figure}[htpb]
\centering
\includegraphics[width=0.73\textwidth]{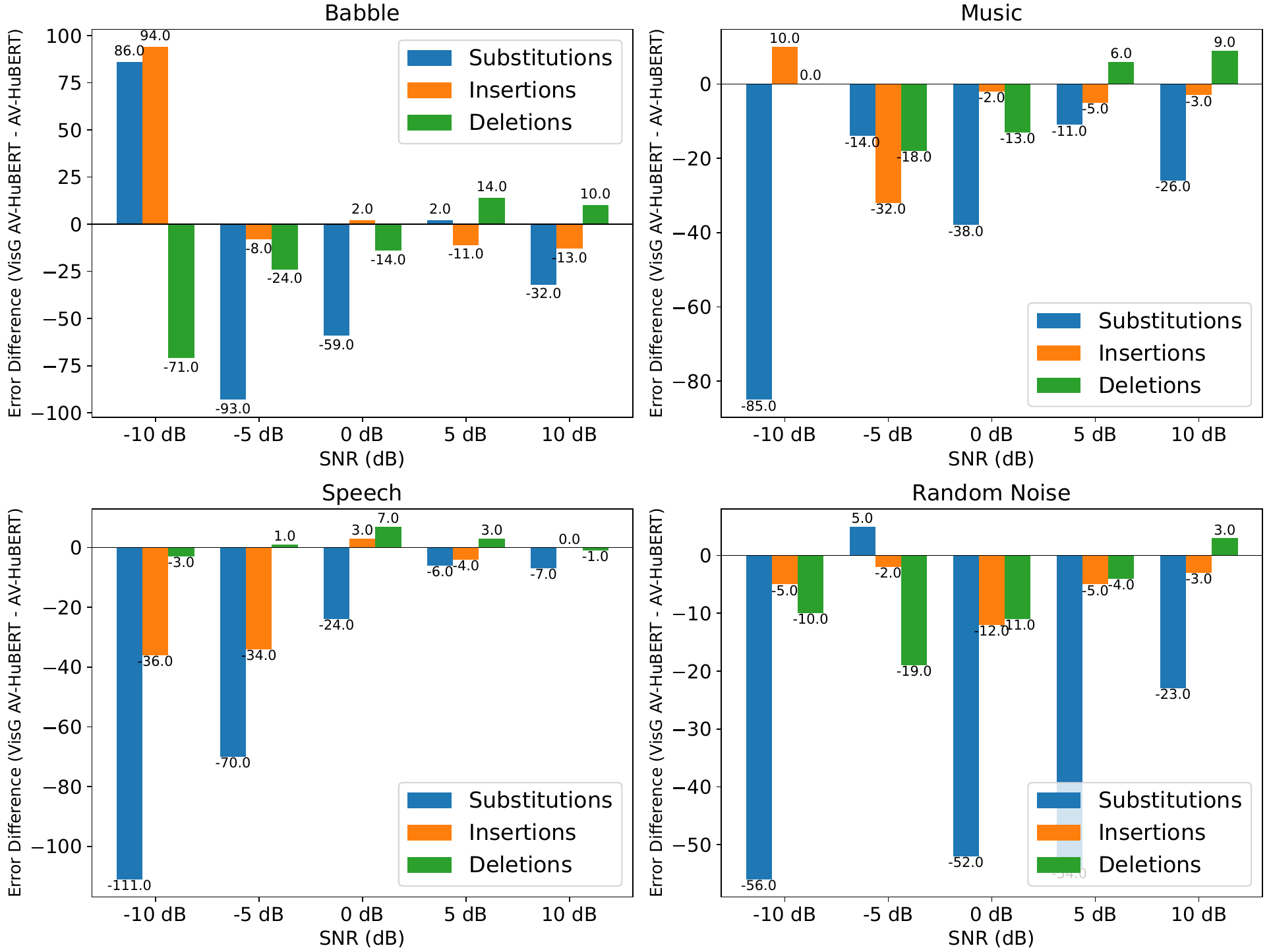}
\caption{Error difference (VisG AV-HuBERT minus AV-HuBERT) by error type across noise conditions and SNR levels (Base model).}
 \label{fig:sids_base}
\end{figure}

In Figure \ref{fig:sids_base}, we visualize the Error Difference, similarly to Section \ref{sec:error_types}. VisG AV-HuBERT demonstrates its strongest impact on substitution errors, particularly under Speech noise, where reductions of 111 errors at -10 dB and 70 errors at -5 dB are observed, accompanied by substantial insertion reductions of 36 and 34 errors respectively. This pattern supports the hypothesis that explicit viseme modelling helps resolve phonetic confusion by providing discriminative articulatory features that constrain the model's predictions. Music noise shows similarly strong substitution reductions at extreme SNR levels (-85 at -10 dB), though with more modest insertion improvements. Random Noise conditions exhibit moderate but consistent substitution reductions across SNR levels (-56 at -10 dB, -52 at 0 dB, -54 at 5 dB), with smaller improvements in insertions and deletions. However, Babble noise reveals a problematic pattern: while substitutions improve at -5 dB (-93) and 0 dB (-59), the model exhibits substantial increases in both substitutions (+86) and insertions (+94) at -10 dB SNR. This error increase explains the WER degradation observed under extreme Babble noise conditions and suggests that when multiple competing speakers are present at very low SNR, the viseme classifier may generate spurious predictions that lead to over-insertion of tokens. Overall, the error analysis confirms that viseme guidance reduces substitution errors through improved articulatory feature learning.

\subsection{Viseme Error Analysis}
\begin{figure}[htpb]
\centering
\includegraphics[width=\textwidth]{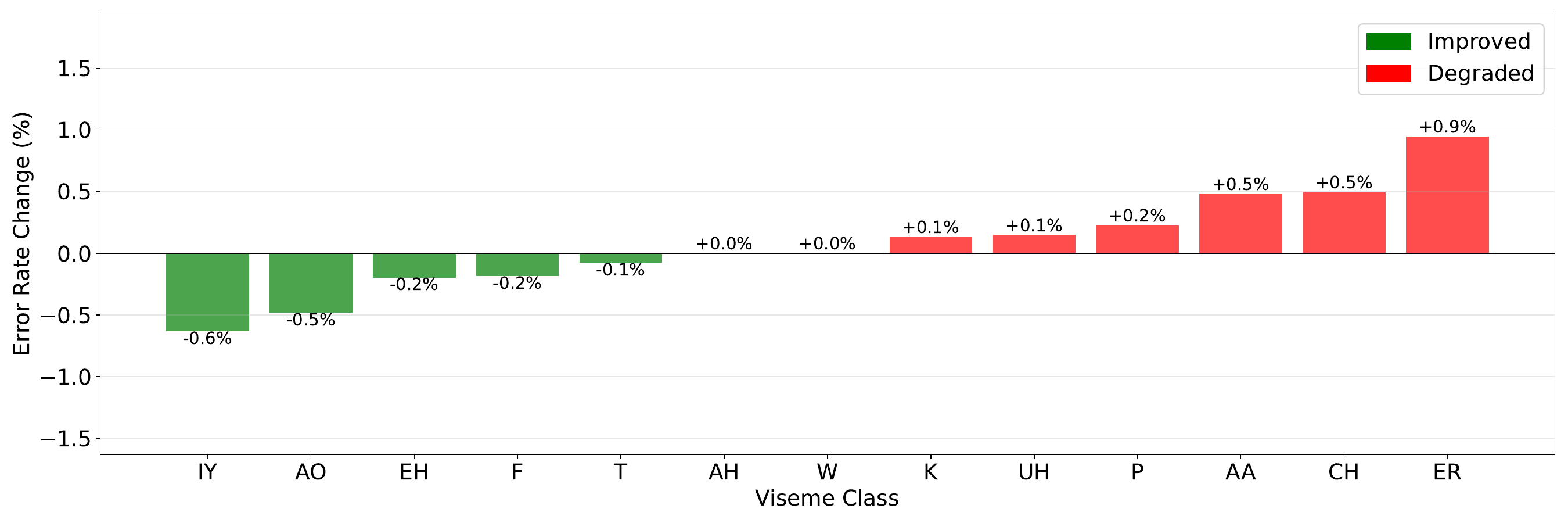}
\caption{Relative VER difference between VisG AV-HuBERT and AV-HuBERT Base.} 
\label{fig:base_ver}
\end{figure}

In this section, we present a thorough investigation of per-viseme error rates and how viseme-guided training affects recognition performance. Figure \ref{fig:base_ver} shows the relative improvement or degradation in per-viseme error rate (VER) for the Base model under clean conditions, computed as VisG AV-HuBERT error rate - AV-HuBERT error rate. Jeffers and Barley \cite{jeffers} provide a visibility ranking, which we use to interpret some of our results.

Analysing the per-viseme error changes in conjunction with visibility rankings reveals an important pattern: the largest improvements occur for less visible vowel visemes (IY: -0.6\%,  AO: -0.5\%, EH: -0.2\%, ), while highly visible visemes show mixed results (ER: +0.9\%,  F: -0.2\%). This suggests that viseme-guided training provides the greatest benefit for phonetic categories where visual information is inherently ambiguous, and audio-visual fusion is most critical for disambiguation. For highly visible visemes, where lip movements are already distinctive, the auxiliary viseme task may introduce competing supervision signals that slightly degrade the learned audio-visual representations. This visibility-dependent performance pattern explains why overall WER improves despite mixed per-viseme results, as the improvements concentrate on visemes where disambiguation is most needed.

\begin{figure}[htpb]
\centering
\includegraphics[width=\textwidth]{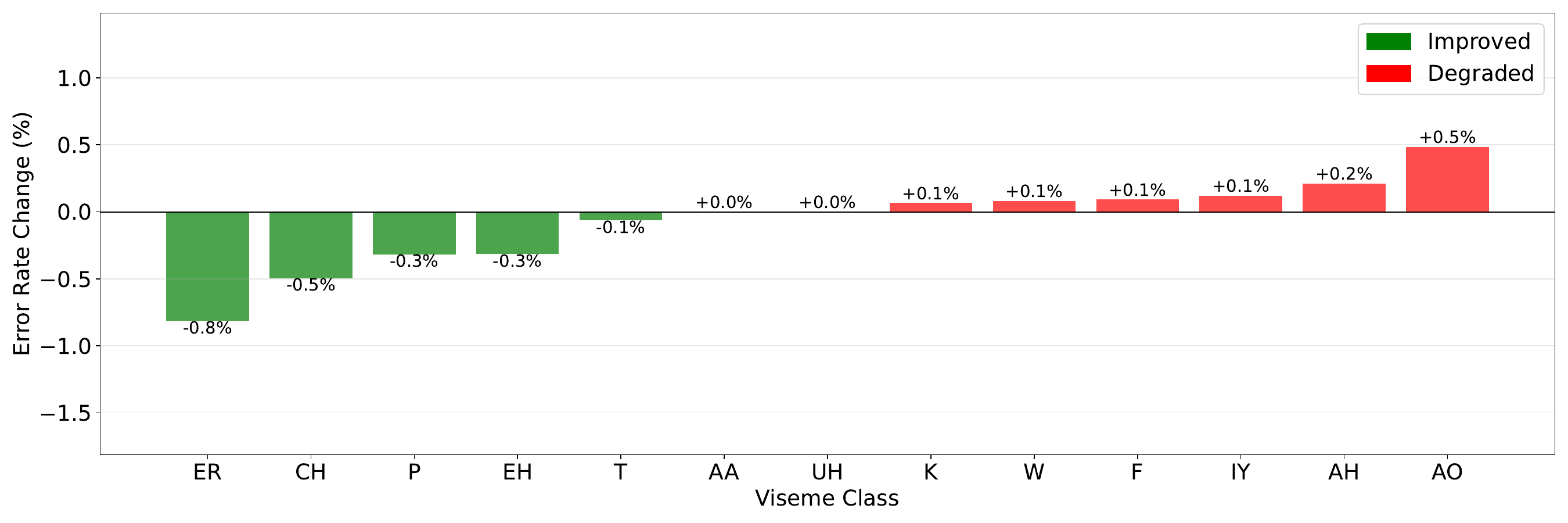}
\caption{Relative VER difference between VisG AV-HuBERT and AV-HuBERT Large.} 
\label{fig:large_ver}
\end{figure}

Figure \ref{fig:large_ver} presents the corresponding per-viseme error rate changes for the Large models.

The figure reveals a striking contrast with the Base model results. While the Base model showed the largest improvements for vowel visemes IY (-0.6\%) and AO (-0.5\%) with degradation for ER (+0.9\%), the Large model exhibits the opposite pattern: ER achieves the strongest improvement (-0.8\%), while IY shows minimal change (+0.1\%) and AO degrades (+0.5\%). Similarly, CH improves substantially for Large (-0.5\%) after degrading for Base (+0.5\%).

This reversal pattern suggests that model capacity fundamentally alters how viseme-guided training affects different articulatory classes. The Large model's greater representational capacity may already capture the visual distinctions for high-frequency vowels (IY, AO) without explicit supervision, making the auxiliary viseme task redundant or even competing for these classes. Conversely, the Large model benefits more from viseme guidance for challenging categories like ER and CH, where the Base model struggled. 

This capacity-dependent pattern reinforces that the auxiliary viseme task serves as an inductive bias rather than direct supervision—its effectiveness depends on what representations the Base model already learns. For smaller models, viseme guidance helps with common vowels; for larger models, it helps disambiguate more complex articulatory patterns.

\subsection{Complexity Results in LRS2}

Following the method detailed in \ref{sentence_complexity}, we train a Language model based on the training and validation texts of LRS2. To illustrate the improvements of viseme guidance, Table \ref{tab:complex_sentences_lrs2} presents representative examples from the LRS2 test set under -10 dB Speech noise, sorted by cross-entropy (CE) score.

\begin{table}[htpb]
\centering
\small
\caption{Test Set Utterances sorted by their CE Score, with predictions from the Large configurations for each model, under -10dB Speech Noise}\label{tab:complex_sentences_lrs2}
\scalebox{0.9}{
\begin{tabular}{|l|l|l|l|}
\hline
\textbf{Ground Truth} &  \textbf{CE} & \textbf{AV-HuBERT} & \textbf{VisG AV-HuBERT}\\
\hline
puerto rican style & 6.55 & border reconcile & borter recon style \\
\hline
squirrel pox virus & 5.47 & we're all boxed virus & we're on post virus \\
\hline
garage and techno funk & 4 & gathering check no funk & garage and techno funk \\
% \hline
% the rochelle to my marvin & 3.91 & reshelled to my marvin & resheld to my mother \\
\hline
great leonard cohen & 3.78 & great lennon cohen & great leonard cohen \\
% \hline
% also known as a tandem & 3.79 & also known as a tandom & are so known as a tandem \\
% \hline
% maybe not uranium & 3.24 & maybe not uranium & maybe not too ricky \\
\hline
causes it no harm & 3.1 & also it's a little harm & causes need to harm \\
\hline
very heavily taxed & 2.65 & very heavily tasks & very heavily taxed \\
% \hline
% the president's chair & 2.24 & president chair & president chair \\
\hline
there's plenty to indulge in & 1.97 & there's plenty to indulging & there's plenty to indulge in \\
\hline
if you've got it & 1.49 & you've got & if you've got it \\
% \hline
% because they've got to & 1.26 & because they've taught to & because they've got to \\
\hline
\end{tabular}
}
\end{table}

\begin{figure}[htpb]
\centering
\includegraphics[width=0.65\textwidth]{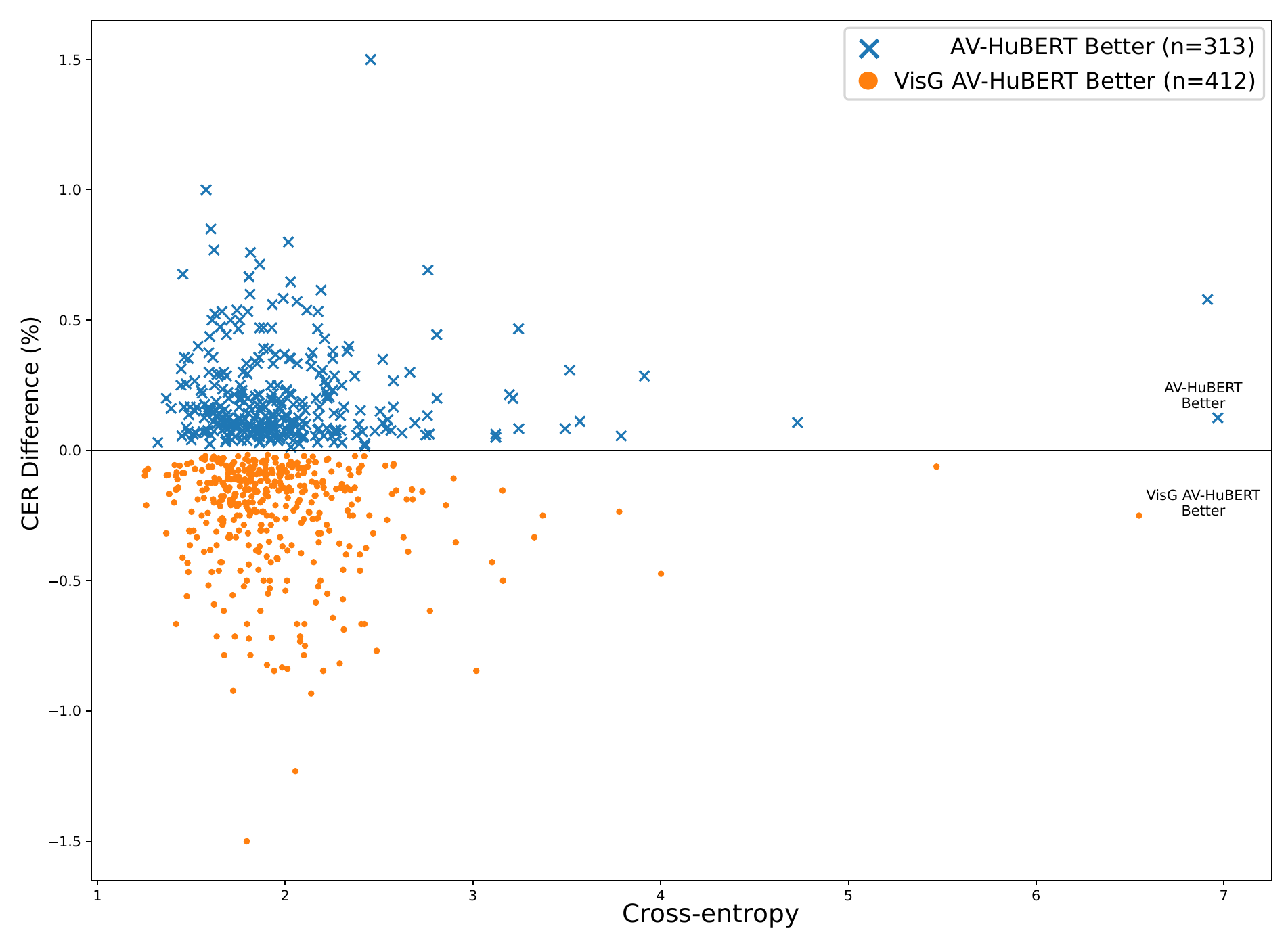}
\caption{CER Difference between VisG AV-HuBERT and AV-HuBERT (Large).}%, under Speech Noise mixed at -10dB.} %  sorted by their complexity
\label{cer_large_lrs2}
\end{figure}

At the highest complexity level, VisG AV-HuBERT generates a prediction that maintains the phonetic structure better, through improved consonant recognition and better word boundary detection. For moderately complex utterances, the proposed model achieves perfect transcription, whereas the baseline produces errors. These improvements stem from viseme guidance helping resolve visually similar phoneme confusions. At lower CE scores, both models perform well, though VisG AV-HuBERT demonstrates consistent advantages in recovering function words. The model correctly predicts "if you've got it" (including the initial "if") and "because they've got to" (distinguishing "got" from "taught"). Notably, the viseme-guided model corrects verb tense in "very heavily taxed" versus the baseline's "tasks," and properly handles gerunds versus infinitives in "there's plenty to indulge in". These qualitative examples demonstrate that auxiliary viseme supervision provides particular benefits for phonetically ambiguous contexts where articulatory visual cues can disambiguate acoustically degraded speech. Figure \ref{cer_large_lrs2} presents the results for the LRS2 test set.

\subsection{Qualitative Examples}
% In this section we provide two examples of utterances. In the first case, the VisG AV-HuBERT improved the utterance WER compared to the baseline. In the second case, VisG AV-HuBERT's performance degraded for this utterance. The examples are taken from VisG AV-HuBERT Large, under Speech noise mixed at -10dB.

\begin{figure}[htbp]
\centering
\includegraphics[width=\textwidth]{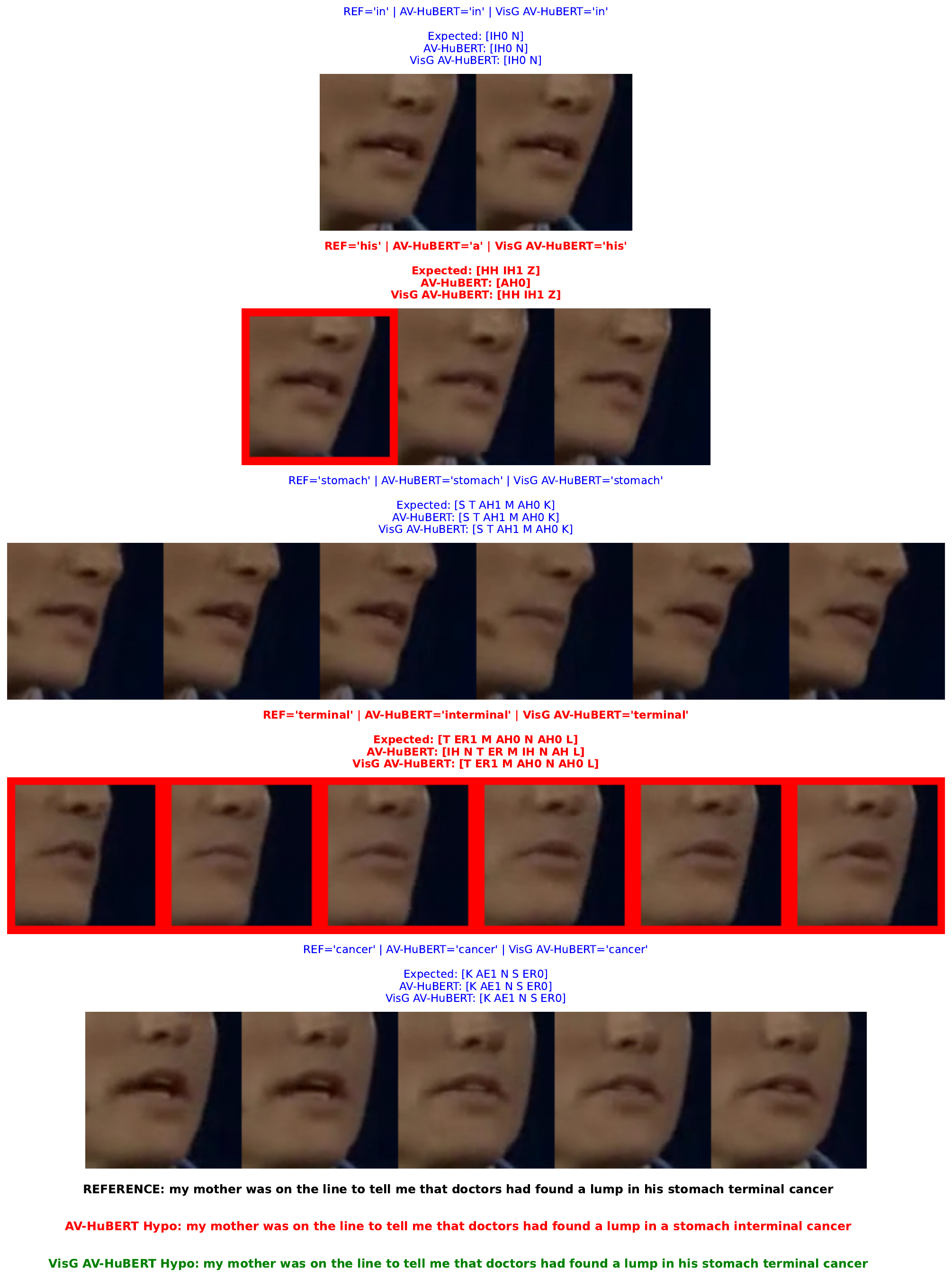}
\caption{Qualitative example showing improved transcription by VisG AV-HuBERT for a Speech noise utterance at -10 dB SNR (LRS3 sample: MvXZzKZ3JYQ\_00001).}
 \label{fig:better_large_speech_-10}
\end{figure}

\begin{figure}[htbp]
\centering
\includegraphics[width=\textwidth]{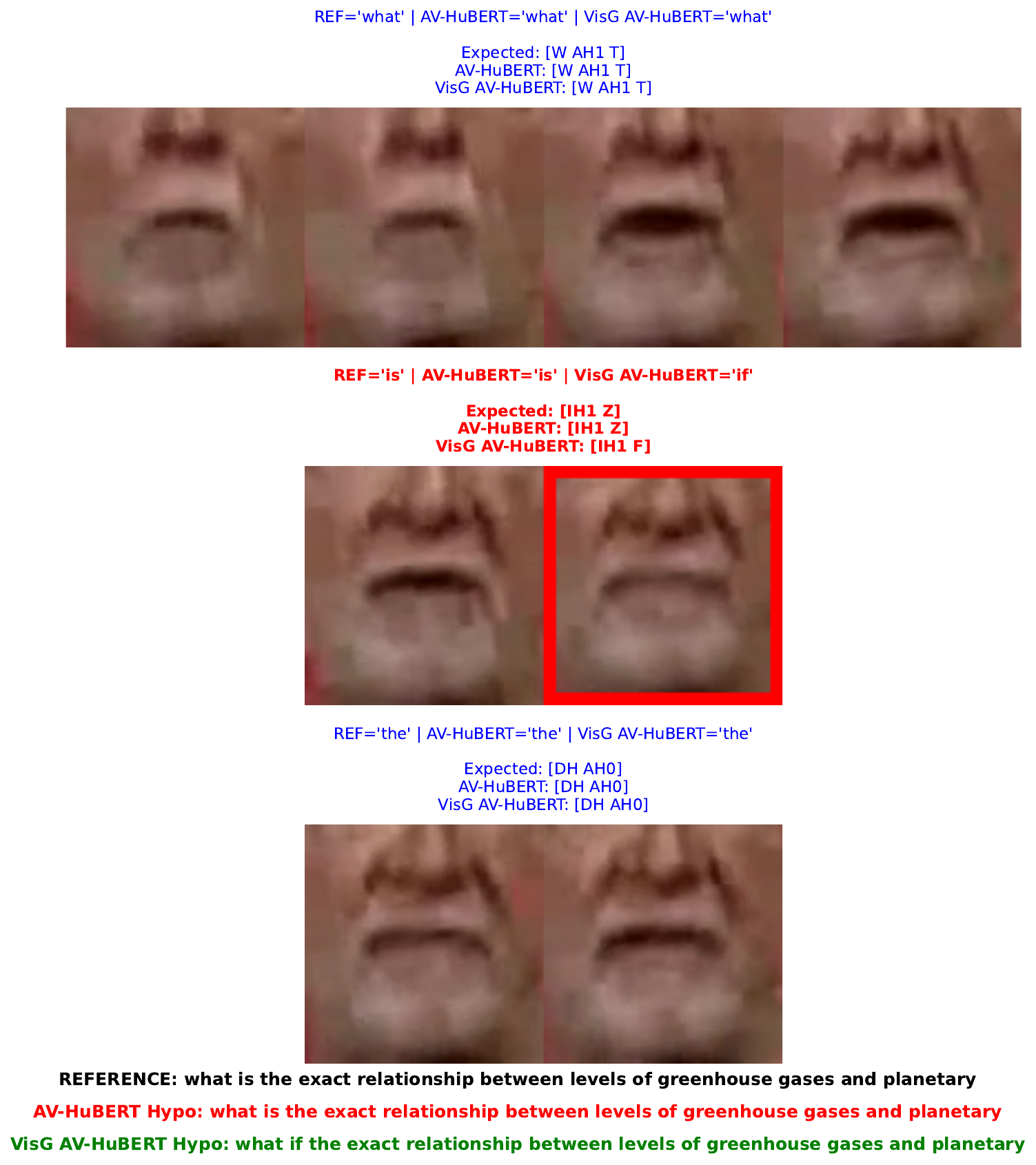}
\caption{Qualitative example showing degraded transcription by VisG AV-HuBERT for a Speech noise utterance at -10 dB SNR (LRS3 sample: SSzRfSJTNW4\_00001).}
 \label{fig:worse_large_speech_-10}
\end{figure}

\end{document}